\documentclass[twocolumn,pra,footinbib,showpacs]{revtex4}

\usepackage{graphicx}
\usepackage{bm}
\usepackage{amsmath}
\usepackage{subfigure}
\usepackage{textcomp}
\usepackage{appendix}
\usepackage{relsize}
\usepackage{rotating}
\usepackage{blkarray}
\usepackage{multirow}

\newcommand{\no}{\noindent}
\newcommand{\non}{\nonumber}
\newcommand{\lb}[1]{\label{#1}}
\newcommand{\beq}[1]{\begin{equation}\label{#1}}
\newcommand{\eeq}{\end{equation}}
\newcommand{\bseq}[1]{\begin{subequations}\label{#1}}
\newcommand{\eseq}{\end{subequations}}
\newcommand{\er}{\eqref}
\newcommand{\plr}[1]{\left(#1\right)}

\newcommand{\bklr}[1]{\left[#1\right]}
\newcommand{\abs}[1]{\left|#1\right|}
\newcommand{\pd}[1]{\partial_{#1}}

\newcommand{\mbf}[1]{\mathbf{#1}}
\newcommand{\ket}[1]{\left|{#1}\right\rangle}
\newcommand{\bra}[1]{\left\langle{#1}\right|}
\newcommand{\eq}{\equiv}

\newcommand{\unit}[1]{\hat{\mathbf{#1}}}
\newcommand{\sg}{\sigma}

\begin{document}

\title{Unified Dynamics of  Electrons and Photons via Zitterbewegung and Spin-Orbit Interaction}

\author{C.C. Leary}
\affiliation{Department of Physics, The College of Wooster, Wooster, OH USA, 44691}
\email{cleary@wooster.edu}
\author{Karl H. Smith}
\affiliation{Department of Physics, The College of Wooster, Wooster, OH USA, 44691}

\date{\today}


\begin{abstract}

We show that when an electron or photon propagates in a cylindrically symmetric waveguide, it experiences both a zitterbewegung effect and a spin-orbit interaction leading to identical propagation dynamics for both particles.  Applying a unified perturbative approach to both particles simultaneously, we find that to first-order in perturbation theory their Hamiltonians each contain identical Darwin (zitterbewegung) and spin-orbit terms, resulting in the unification of their dynamics.  The presence of the zitterbewegung effect may be interpreted physically as the delocalization of the electron on the scale of its Compton wavelength, or the delocalization of the photon on the scale of its wavelength in the waveguide.  The presence of the spin-orbit interaction leads to the prediction of several rotational effects: the spatial or time evolution of either particle's spin/polarization vector is controlled by the sign of its orbital angular momentum quantum number, or conversely, its spatial wave function is controlled by its spin angular momentum. 
\end{abstract}

\maketitle

\section{\label{sec:Intro}Introduction \protect}

It is well-known that when an electron propagates in an inhomogeneous electrostatic potential, its dynamics are influenced by three distinct effects of relativistic origin:  (1) the relativistic mass increase due to the electron's kinetic energy, (2) the delocalization of the electron on the scale of its Compton wavelength corresponding to the so-called zitterbewegung of its motion, and (3) the alteration of the propagation characteristics of the electron arising from the interaction between its spin and orbital angular momenta \cite{Greiner}.  These phenomena were first clearly shown to derive from a common source by Foldy and Wouthuysen, who showed that the Dirac equation for an electron in an external electric field reduces in the nonrelativistic regime to the Schr\"{o}dinger-Pauli equation with three correction terms present in the Hamiltonian.  These terms, often denoted as $\hat{H}_{\text{Rel}}$, $\hat{H}_{\text{Dar}}$, and $\hat{H}_{\text{SO}}$,  correspond respectively with the three aforementioned phenomena, and the latter two may be respectively interpreted as governing interactions between the external electromagnetic field and the electron's delocalized charge distribution and magnetic moment \cite{Foldy49}.  

\begin{figure}
\includegraphics[width=000.475\textwidth]{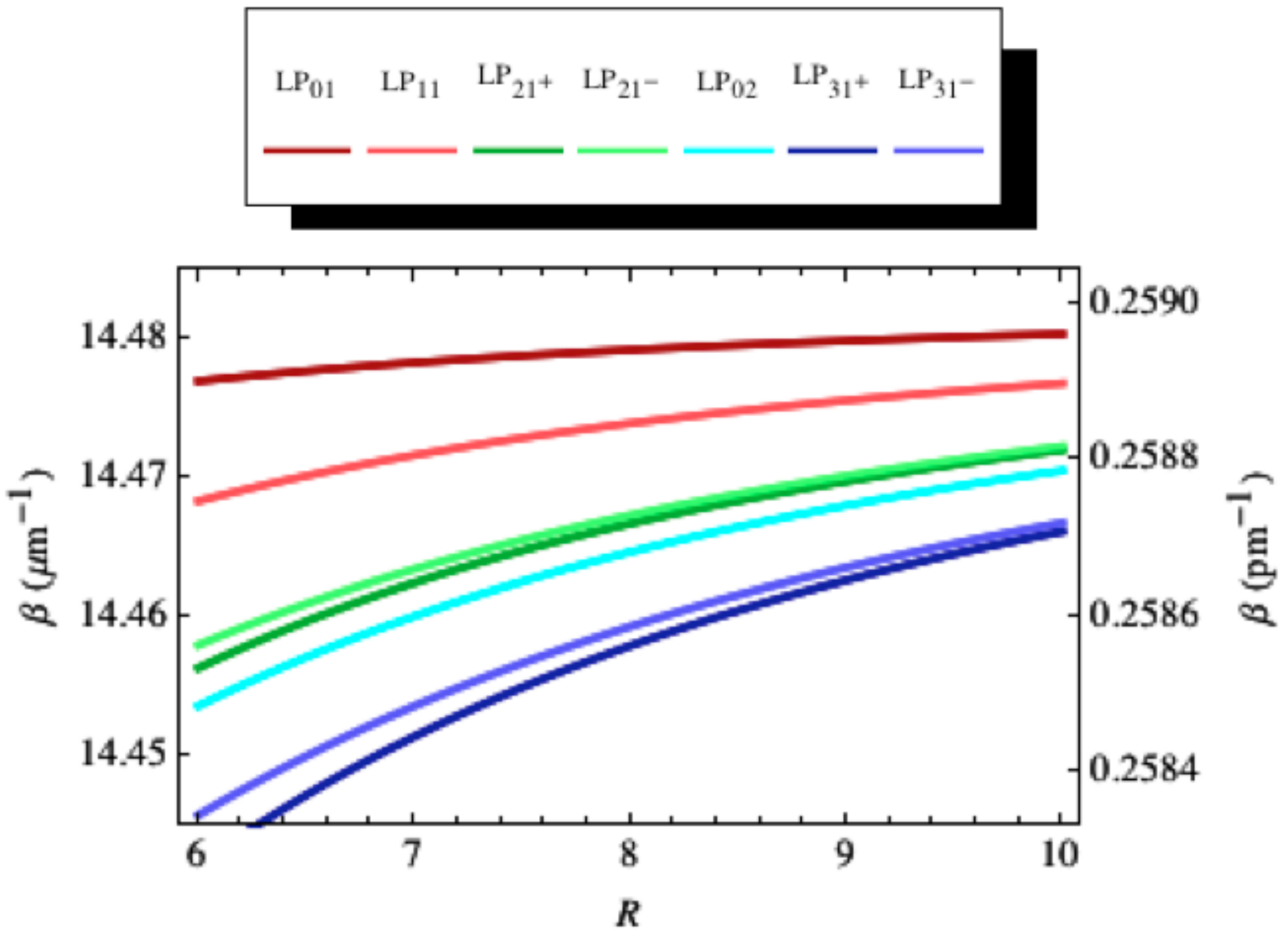}
\vspace{-0.015\textwidth}
\caption{\label{fig:DispRel}  Dispersion curves simultaneously describing an electron or a photon propagating in a cylindrically symmetric waveguide with a step potential/permittivity profile.  Each curve relates the particle's axial momentum (characterized by propagation constant $\beta$) to its normalized frequency  $R$ for a distinct transversely bound state of the system. The left (right) vertical axis delineates  $\beta$ for the photon (electron).  The splitting of certain pairs of dispersion curves due to the spin-orbit interaction is exaggerated by a factor of fifty for purposes of visualization.  In the legend, we have modified the standard optical fiber nomenclature $LP_{\abs{m_{\ell}} n}$ to signify each electronic or photonic bound state (see text for details). }
\end{figure}

In this paper, we show that Maxwell's equations for a photon propagating in a inhomogeneous transparent medium with an axial confining influence may be reduced to a Schr\"{o}dinger-Pauli like equation which includes both Darwin and spin-orbit terms $\hat{H}_{\text{Dar}}$ and $\hat{H}_{\text{SO}}$, in complete analogy with the wave equation for an electron propagating in an axially confining potential.  We find that despite the fact that the photon lacks a charge distribution and a magnetic moment, the Hamiltonians describing the electron and photon dynamics for the zitterbewegung and spin-orbit effects have exactly the same form, with the electron's dimensionless potential energy function $\frac{-eV\plr{r}}{mc^{2}}$ playing the role of the photon's normalized permittivity $\frac{\epsilon\plr{r}}{\epsilon\plr{0}}$, where $r$ denotes the distance from the axis of confinement.  It follows that if the ratio between a particle's wavelength and its effective confinement radius is equal for both particles, the dispersion relations connecting the energy and momentum for each particle's axial momentum eigenstates are given by the same set of curves (see Fig. \ref{fig:DispRel}).

The existence of the photonic spin-orbit interaction was predicted and observed some time ago \cite{Liberman92, Dooghin92}, while the analogous effect for electrons in a cylindrical geometry has more recently been investigated by one of us and others \cite{Bliokh07, Leary08}, as well as a rich connection between the spin-orbit interaction and the geometric Berry phase for both particles (eg. \cite{Berard06, Bliokh08, Leary09}).  A Darwin interaction or zitterbewegung effect has been theoretically proposed for photons propagating in free space \cite{Kobe99} and in anisotropic media \cite{Bliokh07b}, and mentioned in passing for photons in an inhomogeneous, isotropic medium as we study here \cite{Volyar98}.  For electrons, zitterbewegung has been predicted to occur for electron wave packets exhibiting a spin-orbit interaction in condensed matter systems including spintronic semiconductors, graphene, and superconducting systems \cite{Schliemann05, Cserti06}, while the effect has also been experimentally simulated for a one-dimensional Dirac electron via manipulation of a single trapped ion \cite{Gerritsma10}.  Furthermore,  the recent development of experimental techniques for generating electron vortex beams \cite{Uchida10, McMorran11} has brought closer the possibility of observing both the spin-orbit coupling and zitterbewegung effects for electron vortices.  A significant body of work has studied these phenomena  for both particles in the context of semiclassical equations of motion in a trajectory (ray) picture (cf. \cite{Tomita86, Bialynicki-Birula87, Onoda04, Bliokh04, Berard06, Bliokh07b, Bliokh08, Hosten08}).  However, a systematic comparative study of the zitterbewegung and spin-orbit effects for both electrons and photons from the wave equation point of view is lacking in the literature, and is the subject of the present paper.

\section{\label{sec:WaveEQs}Preliminaries\protect}

Our aim here is to compare the wave equations for a monoenergetic electron and photon for the case where each particle experiences a cylindrically symmetric axial confining influence while propagating paraxially with respect to the confinement axis.  Employing cylindrical coordinates $\plr{r,\phi,z}$, we characterize the confining influence by an inhomogeneous potential $V\plr{\rho}$ for electrons, and an inhomogeneous permittivity $\epsilon\plr{\rho}$ for photons, where $\rho\eq\frac{r}{a}$ is a dimensionless radial coordinate, and $a$ is the effective radius of the potential or permittivity.  The requirement of translational invariance in the \textit{z}-direction implies that the monoenergetic wave function describing either particle \footnote{In this work, the transverse electric field  $\Psi\eq\mathbf{E}-\plr{\mathbf{E}\cdot\hat{\mathbf{z}}}\hat{\mathbf{z}}$, which has two nonzero vectorial components, is taken to be the wave function for the photon, as it plays a role analogous to that of the two-component spinorial wave function of the electron throughout.} assumes the following traveling wave form,

\beq{Wave Function}
\Psi=\Psi\left(\rho,\phi\right)e^{i\left(\beta z-\omega  t\right)}
\eeq

\no where $\Psi\left(\rho,\phi\right)$ is a two-component spinor for the case of an electron, and a two-dimensional vector lying in the plane transverse to the confinement axis for a photon.  

  As we will show, under the above conditions the wave equations for both an electron and a photon take the same general form:

\beq{Wave Equation}
\bklr{\mathbf{\nabla}_T^{2} +k^2 \plr{\rho}}\Psi +\hat{H}^{\prime}\Psi= \beta^2 \Psi,
\eeq

\no where $\mathbf{\nabla}_{T}^{2}\eq\mathbf{\nabla}^{2}-\partial_{z}^{2}$ is the transverse Laplacian.  The function $k^2 \plr{\rho}$---which depends implicitly on the particle's angular frequency $\omega$---determines the dispersion relation between $\omega$ and the particle's propagation constant $\beta$.  We find that for a weakly--guided electron or photon, the term $\hat{H}^{\prime}\Psi$ in \er{Wave Equation}---which is defined below for both cases---may be treated as a perturbation of the Helmholtz-type wave equation

\beq{unpert}
 \hat{H}_0 \Psi_0 =\beta_0^2 \Psi_0.
 \eeq 

\no where $ \hat{H}_0 \eq\mathbf{\nabla}_T^{2} +k^2 \plr{\rho}$.
 
  For a given particle energy  $\hbar\omega$, the unperturbed solutions to equation \er{unpert} may be expressed in the form

\begin{equation} \label{Paulistates}
\Psi_0 \eq \ket{n\, m_{\ell}\, \sigma}=\frac{1}{a}\psi_{n\abs{m_{\ell}}}\!\left(\rho \right)e^{im_{\ell } \phi } \unit{e}_{\sg} e^{i\left(\beta _{0} z-\omega t\right)}\end{equation} 

\noindent where $\psi_{n\abs{m_{\ell}}}\left(\rho \right)$ is a dimensionless, real scalar function determined by $k^2\plr{\rho}$.  The wave functions in \er{Paulistates} are expressed via the ket notation in terms of the following quantum numbers, $\{n,\, m_{\ell},\, \sigma \}$, which are respectively associated with the following set of commuting operators, $\{\hat{H}_{0},\,\hat{\ell}_{z},\,\hat{\sg}_{z}\}$, where $\hat{\sg}_{z}\eq\left(\begin{array}{cc} {1}&{0} \\ {0}&{-1} \end{array}\right)$ and $\hat{\ell}_{z}\eq-i\pd{\phi}$ are dimensionless operators  respectively representing the \textit{z}-components of the spin and orbital angular momentum of the particle.  The quantum numbers $n$, $m_{\ell}$, and $\sg$ are constrained to the following integral values: $n=1,2,3, \ldots$; $m_{\ell}=0,\pm1,\pm2, \ldots$; $\sg=\pm1\eq\pm$.  Also in \er{Paulistates}, $\unit{e}_{\sg} \eq\left(\begin{array}{c} {\delta _{\sigma +} } \\ {\delta _{\sigma -} } \end{array}\right)$ is a two-component spinor (for electrons) or a helicty unit vector $\unit{e}_{\sg}\eq\frac{\unit{x}+i\unit{y}}{\sqrt{2}}\delta _{\sigma +} +\frac{\unit{x}-i\unit{y}}{\sqrt{2}}\delta _{\sigma -}$ (for photons) composed of Kronecker delta functions, such that the sign of $\sg$ determines the handedness of the particle's helicity as measured in the laboratory frame. The eigenvalues $\beta_{0}^{2}$ of equation \er{unpert} depend on $n$ and on the absolute value $\abs{m_{\ell}}$, but are independent of $\sg$.  In order to explicitly identify this dependence, we will sometimes denote the squared propagation constant as $\beta_{0}^{2}\eq\plr{\beta_{0}^{2}}_{n\abs{m_{\ell}}}$, or the unsquared one as  $\beta_{0}\eq\plr{\beta_{0}}_{n\abs{m_{\ell}}}$.

  In what follows we derive equation \er{Wave Equation} for both electrons and photons, and discuss the perturbation term $\hat{H}^{\prime}\Psi$ in detail for both cases.  We find that to first-order in perturbation theory, this term---which includes both the spin-orbit interaction and Zitterbewegung effects---assumes effectively identical forms for electrons and photons.

\section{\label{sec:Electron}Electron Wave Equation\protect}

Consider a single monoenergetic electron with mass $m$, charge $-e$, and energy $\hbar\omega$, moving in the presence of an electromagnetic field at a non-relativistic speed with respect to the laboratory frame.  If the electromagnetic field contains sufficiently low (i.e.\ non-relativistic) field energy then the electron may be described by the Dirac equation.  Furthermore, if the electromagnetic field is electrostatic in the laboratory frame, in this frame the magnetic field $\mathbf{B}$ and magnetic vector potential $\mathbf{A}$ may be set to zero, and the electric field $\mathbf{E}$ has zero curl.  In such a case, the Dirac equation for the electron in Gaussian units, as expressed in the Foldy-Wouthuysen representation to order $\left({\tfrac{v}{c}} \right)^{4}$ takes the form \cite{Greiner}

\beq{Dirac}
\hat{H} \Psi=\hbar\omega\Psi
\eeq

\no with

\begin{align} \lb{Foldy} 
&\hat{H} = mc^{2}+\frac{\hat{\mathbf{p}}^{2}}{2m}-eV  \nonumber \\ 
&-\negthinspace\left\{\negthinspace\frac{1}{2m c^{2} } \left(\frac{\hat{\mathbf{p}}^{2}}{2m}\right)^{2}\negthickspace+\frac{e\hbar^2}{8m^{2} c^{2} } \nabla \cdot \mathbf{E}-\frac{e}{2m^{2} c^{2} } \hat{\mathbf{S}}\cdot \left(\mathbf{E}\times \hat{\mathbf{p}}\right)\negthinspace \right\},
\end{align}

\no where $\hat{\mathbf{p}}\eq-i\hbar\nabla$ is the electronic momentum operator, $\hat{\mathbf{S}}\equiv\frac{\hbar}{2}\hat{\mathbf{\sigma}}$ is the spin vector operator of $2\times2$ Pauli matrices, and a standard notation has been used to represent the other electronic properties.

  Assuming a cylindrically symmetric but inhomogeneous potential $V\plr{\rho}$ so that $\mathbf{E}=-\pd{r}V\left(\rho\right)\hat{\mathbf{r}}$, we express $V\plr{\rho}$ in terms of a dimensionless function $\chi\plr{\rho}$ through the relationship 

\beq{NormalizedV}  
W_{\text{e$^-$}}\!\plr{\rho}=W_{\text{e$^-$}}\!\plr{0}+\Delta \chi\plr{\rho}
\eeq

\no where $W_{\text{e$^-$}}\!\plr{\rho}\eq\frac{-eV\plr{\rho}}{mc^{2}}$ is a dimensionless ``normalized'' potential energy, and $\Delta\eq -\plr{W_{\text{e$^-$}}\!\plr{0}-W_{\text{e$^-$}}\!\plr{a}}$. In order to guarantee the existence of transversely bound states, we assume that $\chi$ is zero at the origin and increases monotonically to one at radius $a$, becoming constant thereafter. For simplicity we also fix the arbitrary ``zero point'' potential energy according to $eV\plr{0}=mc^2$, so that the $mc^2$ term may be dropped in the first line of \er{Foldy} and $W\plr{0}$ set to zero in \er{NormalizedV}.  With these substitutions, multiplying both sides of \er{Dirac} by the quantity $-\frac{2m}{\hbar^{2}}$ yields an equation of the form \er{Wave Equation}, with 

\beq{ksqelec}
k^2 \plr{\rho}=\frac{2}{\lambdabar_{C}^2}\plr{\frac{\hbar\omega}{mc^2}-\Delta\chi},
\eeq

\no and 

\beq{EHPrime}
\hat{H}^{\prime}=\hat{H}_{\text{Rel}}^{\text{(e$^-$)}}+\hat{H}_{\text{Dar}}^{\text{(e$^-$)}}+\hat{H}_{\text{SO}}^{\text{(e$^-$)}}, 
\eeq

\no where

\bseq{EPert}
\begin{align}
\hat{H}_{\text{Rel}}^{\text{(e$^-$)}}&=\frac{1}{\lambdabar_{C}^2}\plr{\frac{\hbar\omega}{mc^2}-\Delta\chi}^2, \lb{EPerta} \\
\hat{H}_{\text{Dar}}^{\text{(e$^-$)}}&=-\frac{\Delta}{4a^2}\plr{\chi^{\prime\prime}+\frac{\chi^{\prime}}{\rho}}, \lb{EPertb} \\
\hat{H}_{\text{SO}}^{\text{(e$^-$)}}&=-\frac{\Delta}{a^2}\frac{\chi^{\prime}}{\rho}\,\frac{1}{\hbar}\hat{\mathbf{S}}\cdot \left( \mbox{\boldmath$\rho$}\times \frac{1}{\hbar}\hat{\mathbf{p}}\right) \non  \\
&=-\frac{\Delta}{2a^2}\frac{\chi^{\prime}}{\rho}\bklr{\hat{\sigma}_{z} \hat{\ell}_{z}+\rho\hat{\mathbf{N}}^{\text{($e^{-}$)}}\partial_{z}}, \lb{EPertc}
\end{align}
\eseq

\no with the primes on the $\chi$ functions denoting derivatives with respect to $\rho$. In deriving \er{EPerta}, the zero--order Schr\"{o}dinger relation $\frac{\hat{\mathbf{p}}^2}{2m}=\hbar\omega+eV\plr{\rho}$ has been used, keeping \er{EPerta} accurate to order $\left({\tfrac{v}{c}} \right)^{4}$, and $\lambdabar_{C}\eq\frac{\hbar}{mc}$ is the electron's reduced Compton wavelength.  In \er{EPertc}, the vector dot product involving the spin operator $\hat{\mathbf{S}}$ has been carried out, while $\hat{\mathbf{N}}^{\text{($e^{-}$)}}\eq\hat{\sigma}_{+}\hat{\ell}_{-}-\hat{\sigma}_{-}\hat{\ell}_{+}$, where $\hat{\sigma}_{\pm}$ and $\hat{\ell}_{\pm}$ are the respective raising and lowering operators of the \textit{z}-components of spin and orbital angular momentum for the electronic states $\ket{n\, m_{\ell}\, \sigma}$, such that 

\bseq{mixing}
\begin{align} 
&\hat{\sigma}_{\pm}\hat{\ell}_{\mp}\left|n\,  m_{\ell}\;\,\mp  \right\rangle=\left|n\,  m_{\ell}\!\mp\!1\;\,\pm  \right\rangle, \lb{mixingA} \\
&\hat{\sigma}_{\pm}\hat{\ell}_{\mp}\left|n\,  m_{\ell}\;\,\pm  \right\rangle=0. \lb{mixingB}
\end{align}
\eseq

 The three terms in \er{EHPrime} correspond respectively to the those in curly brackets in \er{Foldy}, and are well-known in the context of the canonical case of a spherically symmetric Coulomb potential, where together they account for all observed spectral phenomena of the hydrogen atom to order $\left({\tfrac{v}{c}} \right)^{4}$ \cite{Shankar}.  In the present cylindrically symmetric case, their interpretation remains the same: $\hat{H}_{\text{Rel}}^{\text{(e$^-$)}}$ arises from the relativistic mass increase due to the electron's kinetic energy, $\hat{H}_{\text{Dar}}^{\text{(e$^-$)}}$ (the Darwin term) accounts for the additional potential energy imparted to the electron due to the so-called zitterbewegung of its motion, and $\hat{H}_{\text{SO}}^{\text{(e$^-$)}}$ gives rise to the electronic spin-orbit interaction. 

\section{\label{sec:Photon}Photon Wave Equation\protect}

Consider now the analogous case of a monoenergetic photon with energy $\hbar\omega$ propagating in a non-magnetic transparent medium with constant permeability $\mu_0$ and cylindrically symmetric but spatially inhomogeneous permittivity $\epsilon\plr{\rho}$.  In this case Maxwell's equations reduce to a single equation which has precisely the same general form \er{Wave Equation} as does the electron wave equation, but with $k^2 \plr{\mbf{\rho}}=\omega^{2}\epsilon\plr{\rho}\mu_{0}$ and $\hat{H}^{\prime}\Psi=\nabla_{T}\bklr{\nabla_{T}\ln\epsilon\left(\rho\right)\cdot\Psi}$ for photons, as shown in \cite{Snyder}.  Here, $\nabla_{T}\eq\nabla-\pd{z}\hat{\mathbf{z}}$ is the transverse gradient, while the transverse part of the electric field $\Psi\eq\mathbf{E}-\plr{\mathbf{E}\cdot\hat{\mathbf{z}}}\hat{\mathbf{z}}$ has the traveling-wave form given in \er{Wave Function} and plays the role of a photonic wave function.  We express $\epsilon\plr{\rho}$ in terms of the dimensionless function $\chi$ introduced previously:

\beq{Permitivity}
W_{\gamma}\plr{\rho}=W_{\gamma}\plr{0}-\Delta\chi\plr{\rho}.
\eeq

\no Here, in analogy to \er{NormalizedV}, $W_{\gamma}\plr{\rho}\eq\frac{\epsilon\plr{\rho}}{\epsilon\plr{0}}\eq\frac{n^{2}\plr{\rho}}{n^{2}\plr{0}}$ is a normalized permittivity  for the medium (or equivalently, its normalized squared refractive index), while $W_{\gamma}\plr{0}=1$ and $\Delta\equiv W_{\gamma}\plr{0}-W_{\gamma}\plr{a}$, so that

\beq{ksqphot}
k^2 \plr{\rho}=\frac{1}{\lambdabar_{\gamma}^2}\plr{1-\Delta\chi},
\eeq

\no in analogy with \er{ksqelec}, where $\lambdabar_{\gamma}\eq\frac{1}{n\plr{0}}\frac{\lambda_{0}}{2\pi}$ is the ``reduced'' photon wavelength at the waveguide center, with $\lambda_{0}$ being the monoenergetic photon's wavelength in vacuum.

In Appendix \ref{App:PhotonPerturbation}, we show that the term $\hat{H}^{\prime}\Psi=\nabla_{T}\bklr{\nabla_{T}\ln\epsilon\left(\rho\right)\cdot\Psi}$ given above may be expressed in the form
 
\beq{PHPrime}
\hat{H}^{\prime}\Psi=\plr{\hat{H}_{\text{Dar}}^{\plr{\gamma}}+\hat{H}_{\text{SO}}^{\plr{\gamma}}}\plr{\mathbf{1}+\mathbf{\hat{N}^{\plr{\gamma}}}}\Psi,
\eeq

\no  with

\bseq{PPert}
\begin{align}
\hat{H}_{\text{Dar}}^{\plr{\gamma}}&=-\frac{1}{1-\Delta\chi}\frac{\Delta}{2a^2}\plr{\chi^{\prime\prime} +\chi' \pd{\rho}+\frac{\chi^{\prime} }{\rho}+\frac{\Delta\plr{\chi^{\prime}}^2}{1-\Delta\chi}}\!, \lb{PPerta} \\
\hat{H}_{\text{SO}}^{\plr{\gamma}}&=-\frac{1}{1-\Delta\chi}\frac{\Delta}{2a^2}\frac{\chi^{\prime}}{\rho}\hat{\sigma}_{z} \hat{\ell}_{z}, \lb{PPertb}
\end{align}
\eseq

\no where $\mathbf{1}$ denotes the identity operator while $\mathbf{\hat{N}^{\plr{\gamma}}}\eq\hat{\sigma}^{2}_{+}\hat{\ell}^{2}_{-}+\hat{\sigma}^{2}_{-}\hat{\ell}^{2}_{+}$, with the angular momentum raising and lowering operators $\hat{\sigma}_{\pm}$ and $\hat{\ell}_{\pm}$ defined in \er{mixing}.

  We stress here that our treatment thus far is exact for photons, in the sense that each solution to equation \er{Wave Equation} with $k^2 \plr{\rho}$ given by \er{ksqphot} and $\hat{H}^{\prime}\Psi$ given by \er{PHPrime}
generates a complete and exact solution $\{\mbf{E,B}\}$ to Maxwell's equations \cite{Snyder}.  We will show that for sufficiently small $\Delta$, the physics described by the Hamiltonians given in \er{EHPrime} and \er{PHPrime} is effectively identical.

\section{\label{sec:Pert}Perturbative Treatment\protect}

If the overall variation of the electronic potential $V\plr{\rho}$ or photonic permittivity $\epsilon\plr{\rho}$ is sufficiently small, the particle is weakly guided, from which it follows that the guided modes are nearly paraxial.  For either particle, these conditions is fulfilled when $\Delta\ll 1$, which suggests the treatment of $\hat{H}^{\prime}$ as a perturbation to equation \er{unpert}.  We are therefore led to calculate the matrix elements of the perturbation $\hat{H}'$ in the unperturbed monoenergetic eigenstates $\ket{n\, m_{\ell}\, \sigma}$, denoted as $\bra{n^{\prime}\, m_{\ell}^{\prime}\, \sigma^{\prime}}\hat{H}^{\prime}\ket{n\, m_{\ell}\, \sigma}$, where the inner product is defined in the position representation as 

\bseq{innerproduct}
\begin{align} 
& \bra{n^{\prime}\, m_{\ell}^{\prime}\, \sigma^{\prime}}\hat{H}^{\prime}\ket{n\, m_{\ell}\, \sigma} \eq  \int{\!\!\!\!\int{\Psi^{\dagger }_{0}\hat{H}^{\prime}\Psi_{0}\,r dr d\phi}} \lb{innerproducta} \\
& =\!\!\int{\!\!\!\!\int{\!\!\psi_{n^{\prime}\abs{m_{\ell}^{\prime}}}e^{-im_{\ell }^{\prime} \phi }\unit{e}_{\sg^{\prime}}^{\dagger}e^{-i\beta _{0} z} \hat{H}^{\prime} \psi_{n\abs{m_{\ell}}}e^{im_{\ell } \phi }\unit{e}_{\sg}e^{i\beta _{0} z} \! \rho d\rho d\phi}}, \lb{innerproductb}
\end{align}
\eseq 

\no In \er{innerproduct}, the dagger superscript denotes the Hermitian conjugate, and the primes on the quantum numbers in the bra vector serve to differentiate them from those in the ket vector, since they are generally distinct.  

We are presently interested in applying perturbation theory to calculate the first-order shifts in $\plr{\beta_{0}^{2}}_{n\abs{m_{\ell}}}$ due to the perturbation $H^{\prime}$, which we denote as $\delta\!\plr{\beta_{0}^{2}}_{n\abs{m_{\ell}}}$.  In order for these first--order shifts to be accurate, it is sufficient that each of the following conditions be met:

\bseq{conditions}
\begin{align}
& \abs{\bra{n\, m_{\ell}\, \sigma}\hat{H}^{\prime}\ket{n\, m_{\ell}\, \sigma}}\;\ll\abs{\plr{\beta_0 ^2}_{n\abs{m_{\ell}}}} \lb{conditionsA} \\
& \abs{\bra{n^{\prime}\, m_{\ell}\, \sigma}\hat{H}^{\prime}\ket{n\, m_{\ell}\, \sigma}}\ll\abs{\plr{\beta_0 ^2}_{n\abs{m_{\ell}}} - \plr{\beta_0 ^2}_{n^{\prime}\abs{m_{\ell}}}\,} \lb{conditionsB} \\
& \abs{\bra{n^{\prime}\, m_{\ell}\, \sigma}\hat{H}^{\prime}\ket{n\, m_{\ell}\, \sigma}\bra{n\, m_{\ell}\, \sigma}\hat{H}^{\prime}\ket{n^{\prime}\, m_{\ell}\, \sigma}} \non \\ 
& \;\;\;\;\;\;\;\;\;\;\;\;\;\;\;\;\;\;\;\;\;\;\;\;\;\;\;\;\;\;\;\;\;\;\; \ll\abs{\bra{n^{\prime}\, m_{\ell}\, \sigma}\hat{H}^{\prime}\ket{n\, m_{\ell}\, \sigma}} \lb{conditionsC}
\end{align}
\eseq

\no where the second and third conditions need only hold for $n^{\prime}\neq n$.  Numerical calculations show that the conditions in \er{conditions} indeed hold in the specific case where $\chi$ has a step profile, for $\Delta\sim0.01$.  As a result of this, we expect \er{conditions} to hold for a wide range of monotonically increasing $\chi$ profiles, and we henceforth assume this to be the case for the profile in question.

In Appendix \ref{App:Elements}, we explicitly derive the matrix elements in \er{innerproduct} for each particle, and show that to first order in perturbation theory, only the diagonal elements contribute to the shifts in $\delta\!\plr{\beta_{0}^{2}}_{n\abs{m_{\ell}}}$, which may be expressed as follows:

\begin{widetext}
\bseq{deg}
\begin{align} 
&\delta\!\plr{\beta_{0}^{2}}_{n\abs{m_{\ell}}}=\bra{n\, m_{\ell}\, \sigma}\hat{H}^{\prime}\ket{n\, m_{\ell}\, \sigma}  =\frac{1}{\lambdabar_{C}^2} \left<\plr{\frac{\hbar\omega}{mc^2}-\Delta\chi}^2\right>_{\!\!n \abs{m_{\ell}}}+\frac{\pi\Delta}{a^2}\plr{\bigg< \chi^{\prime} \pd{\rho} \bigg>_{\!\!n \abs{m_{\ell}}} -\sg m_{\ell }\left< \frac{\chi^{\prime} }{\rho}\right>_{\!\!n \abs{m_{\ell}}}}\!, \non \\
&\text{\textbf{(for electrons)}} \lb{dega} \\
\non \\
&\delta\!\plr{\beta_{0}^{2}}_{n\abs{m_{\ell}}}=\bra{n\, m_{\ell}\, \sigma}\hat{H}^{\prime}\ket{n\, m_{\ell}\, \sigma}   =\;\;\;\;\;\;\;\;\;\;\;\;\;\;\;\;\;\;\;\;\;\;\;\;\;\;\;\;\;\;\;\;\;\;\;\;\;\;\;\;\;\;\;\;\;\;\;\frac{\pi\Delta}{a^2}\plr{\bigg<\chi^{\prime} \pd{\rho} \bigg>_{\!\!n \abs{m_{\ell}}}-\sg m_{\ell }\left< \frac{\chi^{\prime} }{\rho}\right>_{\!\!n \abs{m_{\ell}}}}\!, 
\;\;  \non \\
&\text{\textbf{(for photons)}} \lb{degb}
\end{align}
\eseq 
\end{widetext}

\no where the bracket notation $\left<\hat{O}\right>_{\!\!n \abs{m_{\ell}}}$ denotes the radial integral $\left< \hat{O}\right>_{\!\!n \abs{m_{\ell}}} \eq \int{\psi_{n\abs{m_{\ell}}}\, \hat{O} \; \psi_{n\abs{m_{\ell}}} \rho d\rho} $ for any operator $\hat{O}$ contained within.

Equations \er{deg} are a principal result of this work: the first--order expressions for the propagation constants are identical in form for electrons and photons propagating in a cylindrical waveguide, with the exception of a state-dependent correction associated with the relativistic mass increase of the electron due to its kinetic energy.  However, for an electron with sufficiently small velocity, this term is negligible in comparison with the remaining Darwin and spin--orbit terms, in which case equations \er{dega} and \er{degb} become completely identical in form.  It follows that in the regime of low velocity electrons, the first-order correction $\delta\!\plr{\beta_{0}}_{n\abs{m_{\ell}}}$ to the \emph{un}squared propagation constant, which may be expressed to first order in terms of $\delta\!\plr{\beta_{0}^{2}}_{n\abs{m_{\ell}}}$ by the relation $\delta\!\plr{\beta_{0}}_{n\abs{m_{\ell}}}\approx\frac{1}{2}\frac{\delta(\beta_{0}^{2})_{\!n\!\abs{m_{\ell}}}}{(\beta_0)_{\!n\!\abs{m_{\ell}}}}$, is given for both particles by the following expression:

\bseq{Correction} 
\begin{align}
\delta\!\plr{\beta_{0}}_{n\abs{m_{\ell}}}&=\frac{\pi\Delta}{2 (\beta_0)_{n\!\abs{m_{\ell}}} a^2}\bigg<\chi^{\prime} \pd{\rho} -\sg m_{\ell } \frac{\chi^{\prime} }{\rho}\bigg>_{\!\!n \abs{m_{\ell}}}, \lb{Correctiona} \\
&=\frac{\pi\Delta}{2 \beta_0 a^2}\int{\!\!\chi^{\prime} \psi\plr{\rho\pd{\rho}-\sg m_{\ell}}\psi d\rho}, \lb{Correctionb}
\end{align}
\eseq

\no where $\psi$ and $\beta_0$ are shorthand for $\psi_{n\abs{m_{\ell}}}$ and $(\beta_0)_{n\!\abs{m_{\ell}}}$.

The first-order shifts $\delta\!\plr{\beta_{0}}_{n\abs{m_{\ell}}}$ may therefore be represented for both particles in terms of the following effective Hamiltonian,

\beq{Heff}
\hat{H}_{\text{Eff}}=\frac{\Delta}{4 \beta_0 a^2}\chi^{\prime} \plr{\pd{\rho} - \frac{1}{\rho}\hat{\sigma}_{z} \hat{\ell}_{z}},
\eeq

\no where

\beq{IP}
\delta\!\plr{\beta_{0}}_{n\abs{m_{\ell}}}=\bra{n\, m_{\ell}\, \sigma}\hat{H}_{\text{Eff}}\ket{n\, m_{\ell}\, \sigma}.
\eeq

\no It is remarkable that the simple expressions \er{Correction}--\er{IP} apply to both electrons and photons.

\section{\label{sec:Unified}Unified Particle Dynamics\protect}

The zitterbewegung of the electron or photon associated with the first (Darwin) term in the Hamiltonian \er{Heff} gives rise to a shift in the dispersion curve $\beta\plr{\omega}$ belonging to each eigenstate $\ket{n \, m_{\ell} \, \sg}$.  The magnitude and direction of the shift depends on the choice of $\chi$ profile, such that the zitterbewegung generally gives rise to a larger shift if there are regions where both $\chi$ and $\psi$ are changing rapidly, provided that  $\psi$ has an appreciable magnitude in these regions.  This effect can be interpreted physically as a delocalization of the interaction between the particle and its confining influence: for electrons, the interaction between the propagating electron and the charge distribution associated with the waveguide potential is ``smeared out'' due to the delocalization of the electron on the scale of its Compton wavelength corresponding to its zitterbewegung motion \cite{Foldy49, Shankar}.  The above results suggest that a smearing of the interaction of the propagating photon and the waveguide permittivity occurs due to an analogous delocalization and zitterbewegung of the photon.  As we will show, consideration of the simple case of a step-profile for $\chi$ suggests that $\lambdabar_{\gamma}$, the photon's reduced wavelength at the waveguide center, in this context plays the role of the electron's Compton wavelength as a natural length scale delimiting photon localizability.  These conclusions are supported by a previous prediction of a zitterbewegung motion for a photon in free space with an amplitude on the order of the photoÕ's reduced classical wavelength in vacuum \cite{Kobe99}.

Like the zitterbewegung shift, The spin-orbit interaction associated with the second term in \er{Heff} is also larger if $\chi$ contains regions of rapid variation.  It leads to a splitting of each set of fourfold-degenerate states  $\{\ket{n\, \abs{m_{\ell}}\, +},\;\ket{n\, -\abs{m_{\ell}}\, -},\;\ket{n\, \abs{m_{\ell}}\, -},\;\ket{n\, -\abs{m_{\ell}}\, +}\}$ into two twofold-degenerate pairs, according to the sign of the product of $\sigma m_{\ell}$: $\sigma m_{\ell}>0$ always corresponds to a downward shift for the dispersion curve, and vice versa.  The effect gives rise to several rotational phenomena involving the wave function of either particle, as has been previously discussed by one of us \cite{Leary08, Leary09}.  As an illustration of this, consider the following distinct balanced superpositions of the monoenergetic, unperturbed eigenstates $\ket{n \, m_{\ell} \, \sg}$, 

\begin{subequations} \label{Sup}
\begin{align}
&\frac{1}{\sqrt{2}}\Big(\ket{n \, m_{\ell} \, \sg}+\ket{n \, m_{\ell} \, -\!\sg}\Big)\propto e^{im_{\ell}\phi}\left(\hat{\mathbf{e}}_{\sigma}+\hat{\mathbf{e}}_{-\sigma}\right), \label{Sup_a} \\
&\frac{1}{\sqrt{2}}\Big(\ket{n \, m_{\ell} \, \sg}+\ket{n \, -\!m_{\ell} \,\, \sg}\Big)\propto \cos\left(\left|m_{\ell}\right|\phi\right)\hat{\mathbf{e}}_{\sigma}. \label{Sup_b}
\end{align}
\end{subequations}

\begin{figure}
\includegraphics[width=000.475\textwidth]{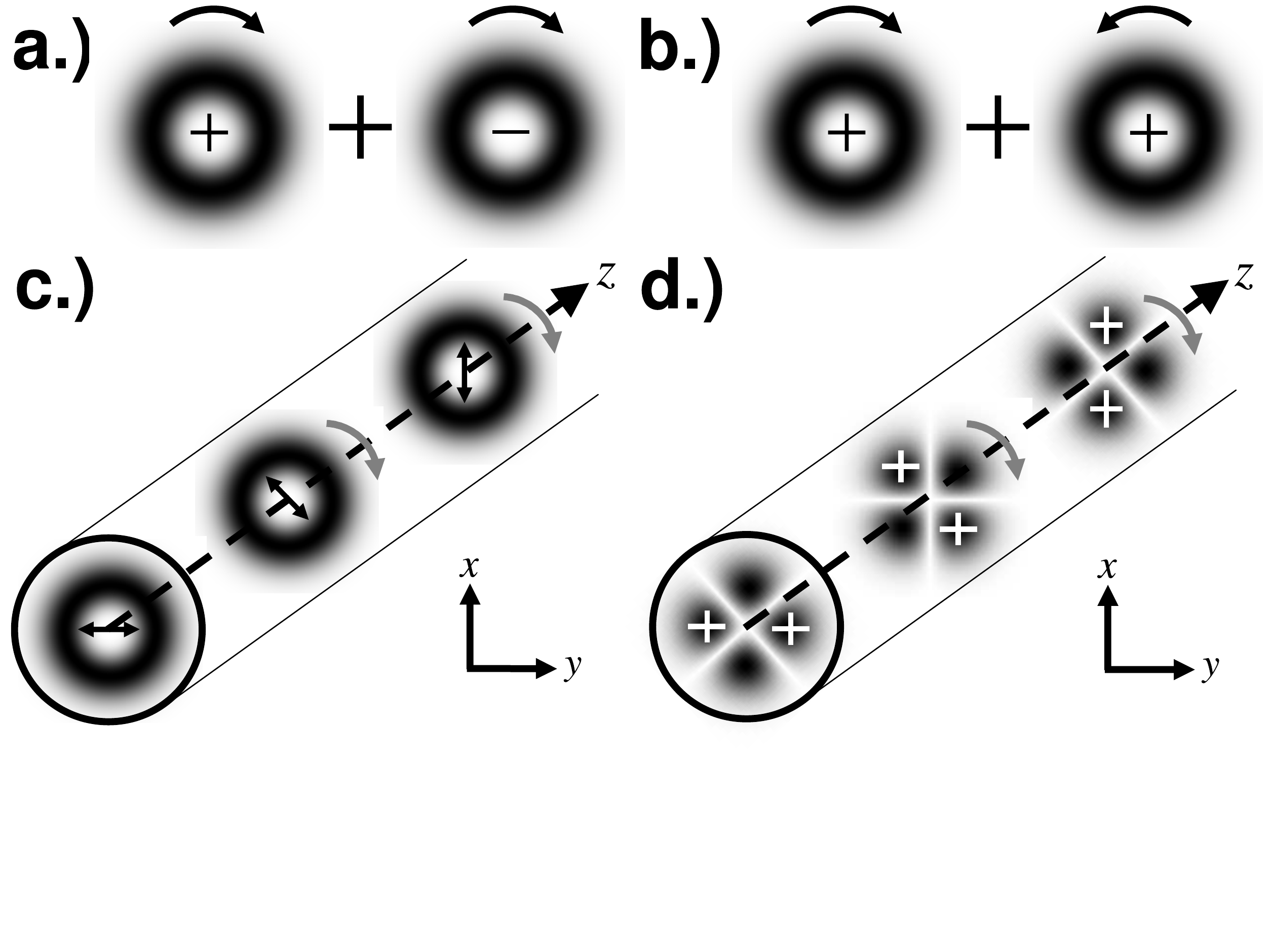}
\caption{\label{fig:SOI} (a) An OAM eigenstate with $\left|m_{\ell}\right|=2$ in a balanced superposition of $+$ and $-$ SAM states (see equation \eqref{Sup_a}). The $\pm$ signs contained within the transverse spatial profiles indicate the SAM of the contributing state, while the arrows indicate its OAM handedness. (b) A SAM eigenstate in a balanced superposition of right and left-handed OAM states with $\left|m_{\ell}\right|=2$ (see equation \eqref{Sup_b}). When states (a) and (b) propagate down a straight waveguide, the spin (polarization) vector of the state in (a) (see equation \eqref{Evo_a}) and the transverse spatial profile of the state in (b) (see equation \eqref{Evo_b}) exhibit azimuthal rotation, as shown in (c) and (d), with the sense of rotation controlled by the sign of the OAM and SAM quantum numbers, respectively. The straight arrows in (c) denote the orientation of the state's spin (polarization) vector, while the white plus signs in (d) represent relative transverse phase.}
\end{figure}

\no  The above wave functions each represent a superposition between a state with \textit{parallel} spin angular momentum (SAM) and orbital angular momentum (OAM) vectors (that is, the product $\sigma m_{\ell}$ is positive) and a state with \textit{anti}-parallel SAM and OAM ($\sigma m_{\ell}$ negative).  They are represented pictorially in Figs. \ref{fig:SOI}(a) and \ref{fig:SOI}(b) for the case where $\abs{m_{\ell}}=2$.

According to \er{Correction}, upon traveling a distance $z$ along the waveguide, a particle in the state $\ket{n \, m_{\ell} \, \sg}$ picks up a phase factor of the following form:

\beq{shift}
\ket{n \, m_{\ell} \, \sg} \to \ket{n \, m_{\ell} \, \sg}e^{i\plr{\abs{\delta\beta_{\text{D}}}-\sg\mu\abs{\delta\beta_{\text{SO}}}} z},
\eeq

\no where $\delta\beta_{\text{D}}\eq\frac{\pi\Delta}{2 \beta_0 a^2}\!\left<\chi^{\prime} \pd{\rho} \right>_{\!n \abs{m_{\ell}}}$ denotes the Darwin shift, $\delta\beta_{\text{SO}}\eq \frac{\pi\Delta}{2 \beta_0 a^2} \left< \frac{\chi^{\prime} }{\rho}\right>_{\!\!n \abs{m_{\ell}}}$ denotes the spin--orbit correction, and $\mu\eq\frac{m_{\ell}}{\abs{m_{\ell}}}$ denotes the absolute sign of the OAM quantum number.  As a result of this phase accumulation, the respective states \eqref{Sup_a} and \eqref{Sup_b} evolve into 

\begin{subequations} \label{Evo}
\begin{align}
\frac{e^{i\abs{\delta\beta_{\text{D}}}z}}{\sqrt{2}}&\Big(\ket{n \, m_{\ell} \, \sg} e^{-i\sigma\mu\abs{\delta\beta_{\text{SO}} z}}+\ket{n \, m_{\ell} \, -\!\sg} e^{+i\sigma\mu\abs{\delta\beta_{\text{SO}}} z}\Big) \nonumber \\
&\!\!\!\!\!\!\!\!\! \propto \psi_{n \left|m_{\ell}\right|}\left(\rho\right)e^{i m_{\ell}\phi}\left(\hat{\mathbf{e}}_{+}e^{-i\mu\left|\delta\beta_{\text{SO}}\right| z}+\hat{\mathbf{e}}_{-}e^{+i\mu\left|\delta\beta_{\text{SO}}\right| z}\right)\!,
\label{Evo_a} \\
\frac{e^{i\abs{\delta\beta_{\text{D}}}z}}{\sqrt{2}}&\Big(\ket{n \, m_{\ell} \, \sg} e^{-i\sigma\mu\abs{\delta\beta_{\text{SO}}} z}+\ket{n \, -\!m_{\ell} \,\, \sg} e^{+i\sigma\mu\abs{\delta\beta_{\text{SO}}} z}\Big) \nonumber \\
& \propto \psi_{n \left|m_{\ell}\right|}\left(\rho\right)\cos\left(\left|m_{\ell}\right|\phi-\sigma\left|\delta\beta_{\text{SO}}\right| z\right) \hat{\mathbf{e}}_{\sigma}. \label{Evo_b} 
\end{align}
\end{subequations}

 For a monoenergetic photon, \eqref{Evo_a} describes a linearly polarized OAM eigenstate whose polarization vector rotates with increasing $z$ as shown in Fig.\ \ref{fig:SOI}(c), such that in a Cartesian basis it can be written as $\cos\left(\left|\delta\beta_{\text{SO}}\right| z\right)\hat{\mathbf{x}}+\mu\sin\left(\left|\delta\beta_{\text{SO}}\right| z\right)\hat{\mathbf{y}}$. Similarly, the expectation value of a monoenergetic electron's spin vector, which rotates in an analogous manner, is $\langle\hat{\mathbf{S}}\rangle=\tfrac{\hbar}{2}\left[\cos\left(2\left|\delta\beta_{\text{SO}}\right| z\right)\hat{\mathbf{x}}+\mu\sin\left(2\left|\delta\beta_{\text{SO}}\right| z\right)\hat{\mathbf{y}}\right]$.  In contrast to \eqref{Evo_a}, \eqref{Evo_b} describes a SAM eigenstate with a rotating orbital state, which has the same form for both particles (see Fig.\ \ref{fig:SOI}(d)). These effects may be viewed as a spatial beating between two waves of identical frequency $\omega$ but with slightly different propagation constants $\beta_{0}\pm\left|\delta\beta_{\text{SO}}\right|$ which have been split by the spin-orbit interaction.  More generally, initial two--state superpositions of the general form
 
\bseq{gensuper}
\begin{align}
&\cos{\plr{\frac{\Theta}{2}}}e^{-i\frac{\Phi}{2}}\ket{n \, m_{\ell} \, \sg}+ \sin{\plr{\frac{\Theta}{2}}}e^{+i\frac{\Phi}{2}}\ket{n \, m_{\ell} \, -\!\sg}, \lb{gensupera} \\
&\cos{\plr{\frac{\Theta}{2}}}e^{-i\frac{\Phi}{2}}\ket{n \, m_{\ell} \, \sg}+ \sin{\plr{\frac{\Theta}{2}}}e^{+i\frac{\Phi}{2}}\ket{n \, -\!m_{\ell} \,\, \sg},  \lb{gensuperb}
\end{align}
\eseq

\no where $\Theta$ and $\Phi$ are the spherical polar angles of the Bloch vector in the relevant two--state space (either SAM or OAM), undergo a precession of the Bloch vector in the azimuthal $\plr{\Phi}$ direction at an angle $\Theta$ due to the spin--orbit shift contained in \er{shift}, with a precession rate of $\abs{\delta\beta_{\text{SO}}}$ per unit $z$.  

In a similar way, each of the spin-orbit interaction effects described above may also occur as a function of time.  Consider the transformation of the eigenstates in \er{gensupera} and \er{gensuperb} according to 

\bseq{trans}
\begin{align}
&\ket{n \, m_{\ell} \, \pm\sg}\to\ket{n \, m_{\ell} \, \pm\sg}e^{i\mu\plr{\abs{\delta\beta_{\text{SO}}}z-\abs{\delta\omega}}t}, \lb{transa} \\
&\ket{n \, \pm m_{\ell} \, \sg}\to\ket{n \, \pm m_{\ell} \, \sg}e^{i\sg\plr{\abs{\delta\beta_{\text{SO}}}z-\abs{\delta\omega}}t}, \lb{transb} 
\end{align}
\eseq

\no respectively, where $\abs{\delta\omega}\eq\frac{\omega}{\beta_{0}}\abs{\delta\beta_{\text{SO}}}$ such that the transformed eigenstates remain eigenstates of \er{unpert}.  Under this transformation, each of the two--state superpositions in \er{gensuper} undergo azimuthal SAM or OAM Bloch vector precession in time with a precession rate of $\abs{\delta\omega}$ per unit $t$ due to the spin--orbit shift contained in \er{shift}.  Consequently, for balanced, in--phase initial superpositions where $\plr{\Theta,\Phi}=\plr{\frac{\pi}{2},0}$, the results \er{Evo} hold under the transformations \er{trans} of the respective equations \er{Sup}, provided that the substitution $\abs{\delta\beta_{\text{SO}}}z\to\abs{\delta\omega}t$ is made in equations \er{Evo}.  Since the direction of Bloch vector precession in spin/polarization space described in \er{Evo_a} is determined by the absolute sign of the OAM quantum number $m_{\ell}$, this effect can be thought of as an orbit--controlled rotation of the particle spin in space or in time.  Conversely, the Bloch vector precession in OAM space of \er{Evo_b} is determined by the absolute sign of the SAM quantum number $\sg$, and one may think of this effect as spin-controlled orbital rotation.

\section{\label{sec:Step}Example: Step Profile\protect}

As a concrete application of the result \er{Correction}, consider the case of a step profile at the boundary, such that $\chi\plr{\rho}=\theta\!\plr{\rho-1}$ and $\chi^{\prime}\plr{\rho}=\delta\!\plr{\rho-1}$, where $\theta$ and $\delta$ are the Heaviside step and Dirac delta functions, respectively.  In this case, $k^2\plr{\rho}$ becomes piecewise constant, so that the unperturbed wave equation \er{unpert} reduces to Bessel's equation, with solutions of the form \er{Paulistates}, with 

\bseq{StepSols}
\begin{align}
\frac{1}{N}\psi_{n\abs{m_{\ell}}}\left(\rho \right)&=J_{\abs{m_{\ell}}}\plr{\kappa_{0} a \rho} \;\;\;\;\;\;\;\;\;\; \text{for} \;\;\;\;\;\rho\leq 1, \lb{StepSolsa} \\
\frac{1}{N}\psi_{n\abs{m_{\ell}}}\left(\rho \right)&=K_{\abs{m_{\ell}}}\plr{\tilde{\kappa}_{0} a \rho} \;\;\;\;\;\;\;\;\; \text{for} \;\;\;\;\;\rho\geq 1,  \lb{StepSolsb}
\end{align}
\eseq

\no where $J_{\abs{m_{\ell}}}\plr{\kappa_{0} a \rho}$ is a Bessel function of the first kind, while $K_{\abs{m_{\ell}}}\plr{\tilde{\kappa}_{0} a \rho}$ is a modified Bessel function of the second kind, and 

\bseq{kappas}
\begin{align}
& \kappa_0\eq \sqrt{k^2\plr{0}-\plr{\beta_{0}^{2}}_{n\abs{m_{\ell}}}} \lb{kappasa} \\
& \tilde{\kappa}_{0}\eq\sqrt{k^2\plr{a}-\plr{\beta_{0}^{2}}_{n\abs{m_{\ell}}}} \lb{kappasb}
\end{align}
\eseq

\no (both of which depend on $n$ and $\abs{m_{\ell}}$) are the respective transverse wave numbers of the particle wave function inside and outside the boundary.  Also in \er{StepSols}, $N$ is a normalization factor given by 

\begin{align} \label{Norm} 
N&\eq\bklr{\int_{0}^{\infty}{\!\!\frac{1}{N^{2}}\psi_{n \, \abs{m_{\ell}}}^2\plr{\rho}} \rho d\rho }^{-\frac{1}{2}}\non \\
&=\frac{1}{\sqrt{\pi}} \frac{1}{J_{\left|m_{\ell } \right|}\left(\kappa _{0} a\right)} \left\{\frac{K_{\left|m_{\ell } \right|-1} \left(\tilde{\kappa }_{0} a\right)K_{\left|m_{\ell } \right|+1} \left(\tilde{\kappa }_{0} a\right)}{K_{\left|m_{\ell } \right|}^{2} \left(\tilde{\kappa }_{0} a\right)} \right. \nonumber \\ 
& \;\;\;\;\;\;\;\;\;\; \;\;\;\;\;\;\;\;\;\; \;\;\;\;\;\;\;\; \left.-\frac{J_{\left|m_{\ell } \right|-1} \left(\kappa _{0} a\right)J_{\left|m_{\ell } \right|+1} \left(\kappa _{0} a\right)}{J_{\left|m_{\ell } \right|}^{2} \left(\kappa _{0} a\right)} \right\}^{-\frac{1}{2}}.
\end{align} 

Application of the appropriate boundary conditions (Schr\"{o}dinger/Dirac \cite{Leary08} or Maxwell \cite{Snyder}) on the electron or photon wave function given in \er{StepSols} results in the following (unperturbed) characteristic equation for either particle:

\beq{char}
\kappa_{0} a\frac{J_{\left|m_{\ell } \right|+1} \left(\kappa_{0} a\right)}{J_{\left|m_{\ell } \right|} \left(\kappa_{0} a\right)} \! = \! \sqrt{R^{2}-\kappa_{0}^{2} a^{2}}\frac{K_{\left|m_{\ell } \right|+1} \plr{\!\sqrt{R^{2}-\kappa_{0}^{2} a^{2}}}}{K_{\left|m_{\ell } \right|} \plr{\!\sqrt{R^{2}-\kappa_{0}^{2} a^{2}}}},
\eeq

\no where

\bseq{Rparam}
\begin{align}
R_{e}&\eq\frac{a_{e}}{\lambdabar_{\text{C}}}\sqrt{2\Delta_{e}} \;\;\;\;\;\;\;\;\;\; \text{\textbf{(for electrons)}}, \lb{Rparama} \\
R_{\gamma}&\eq\frac{a_{\gamma}}{\lambdabar_{\gamma}}\sqrt{\Delta_{\gamma}} \;\;\;\;\;\;\;\;\;\;\;\, \text{\textbf{(for photons)}}, \lb{Rparamb}
\end{align}
\eseq

\no In \er{Rparam},  $R_{e}$, $a_{e}$, and $\Delta_{e}$, are respectively the waveguide parameter, waveguide radius, and step strength parameter of the electron waveguide, while $R_{\gamma}$, $a_{\gamma}$, and $\Delta_{\gamma}$ denote the same quantities for the photon case.  We will continue to denote each of these parameters by their unsubscripted forms $R$, $a$, and $\Delta$ when describing both cases simultaneously.

  For given values of $R$, $n$, and $m_{\ell}$, we solve equation \er{char} numerically for $\kappa_{0}$, which yields $\tilde{\kappa }_{0}$ through equations \er{kappas}, thereby explicitly determining the wave function $\psi$ in \er{StepSols}.  We then use our knowledge of $\psi$ and  equation \er{Correction} to calculate $\delta\!\plr{\beta_{0}}_{n\abs{m_{\ell}}}$, which yields

\begin{widetext}
\beq{dbstep}
\delta\!\plr{\beta_{0}}_{n\abs{m_{\ell}}}=\frac{\Delta}{2\beta_0 a^2}\plr{\frac{\plr{\kappa_0 a} J'_{\abs{m_{\ell}}}\plr{\kappa_{0} a}}{J_{\abs{m_{\ell}}}\plr{\kappa_{0} a}}-\sigma m_{\ell}}\left\{\frac{K_{\left|m_{\ell } \right|-1} \left(\tilde{\kappa }_{0} a\right)K_{\left|m_{\ell } \right|+1} \left(\tilde{\kappa }_{0} a\right)}{K_{\left|m_{\ell } \right|}^{2} \left(\tilde{\kappa }_{0} a\right)}-\frac{J_{\left|m_{\ell } \right|-1} \left(\kappa _{0} a\right)J_{\left|m_{\ell } \right|+1} \left(\kappa _{0} a\right)}{J_{\left|m_{\ell } \right|}^{2} \left(\kappa _{0} a\right)} \right\}^{-1} . 
\eeq
\end{widetext}

\no where the prime on the Bessel function denotes a derivative with respect to its argument: $J'_{\abs{m_{\ell}}}\plr{\kappa_{0} a}\eq\frac{\partial}{\partial\plr{\kappa_{0}a\rho}}J_{\abs{m_{\ell}}}\plr{\kappa_{0} a \rho}\Big|_{\rho=1}$.  From equations \er{ksqelec}, \er{ksqphot}, and \er{kappasa}, we have the following relations for the electronic and photonic unperturbed propagation constants $\plr{\beta_{0}^{e}}_{n\abs{m_{\ell}}}$ and $\plr{\beta_{0}^{\gamma}}_{n\abs{m_{\ell}}}$:

\bseq{beta0}
\begin{align}
\plr{\beta_{0}^{e}}_{n\abs{m_{\ell}}}&=\frac{1}{a_{e}}\sqrt{\plr{\frac{a_{e}}{\lambdabar_{\text{dB}}}}^{2}-\plr{\kappa_0 a_{e}}^{2}} \;\;\;\; \text{\textbf{(for electrons)}}, \lb{beta0a} \\
\plr{\beta_{0}^{\gamma}}_{n\abs{m_{\ell}}}&=\frac{1}{a_{\gamma}}\sqrt{\plr{\frac{a_{\gamma}}{\lambdabar_{\gamma}}}^{2}-\plr{\kappa_0 a_{\gamma}}^{2}} \;\;\;\;\;\, \text{\textbf{(for photons)}}, \lb{beta0b} 
\end{align}
\eseq

\no where $\lambdabar_{\text{dB}}=\sqrt{\frac{\hbar}{2 m \omega}}$ is the electron's (reduced) deBroglie wavelength.  In order to more directly compare the electron and photon dynamics for the step profile case, we fix the ratio of waveguide radius to particle momentum in \er{beta0} for both particles according to 

\beq{fixbeta}
\frac{a_{e}}{\lambdabar_{dB}}=\frac{a_{\gamma}}{\lambdabar_{\gamma}}.
\eeq

\no by which it follows that $\plr{\beta_{0}^{e}}_{n\abs{m_{\ell}}}=\frac{a_{\gamma}}{a_{e}}\plr{\beta_{0}^{\gamma}}_{n\abs{m_{\ell}}}$, since $\kappa_{0} a$ has the same value for both particles. Additionally, we fix $R_{e}=R_{\gamma}$ in equation \er{Rparam} by requiring that 

\beq{fixR}
\Delta_{e}=\frac{1}{2}\plr{\frac{\lambda_{C}}{\lambda_{dB}}}^{2}\Delta_{\gamma}.
\eeq

In Fig. \ref{fig:DispRel} we have plotted---simultaneously for both electrons and photons---the corrected propagation constant $\beta_{n\abs{m_{\ell}}}\eq\plr{\beta_{0}}_{n\abs{m_{\ell}}}+\delta\!\plr{\beta_{0}}_{n\abs{m_{\ell}}}$ \textit{vs}. $R$ for eigenstates with the first few allowed values of $n$ and $\abs{m_{\ell}}$, subject to the constraints \er{fixbeta} and \er{fixR}.  In the legend, we have used the standard optical optical fiber nomenclature $LP_{\abs{m_{\ell}} n}$ to describe the state of an electron or photon \cite{Saleh}, with the following addition: $LP_{\abs{m_{\ell}} n^{\pm}}$ denotes a particle wave function $\ket{n\, m_{\ell}\, \sigma}$ such that sign of the product $\sg m_{\ell}$ is positive or negative as indicated.  Parameters for the photon case were chosen for a helium neon laser propagating in a commercially available step-index optical fiber, where $n\plr{0}=1.46$, $\Delta=0.014$, and $\lambdabar_{\gamma}=\frac{1}{n\plr{0}}\frac{632.8}{2 \pi}$ nm.  Conversely, we chose the electron to have a de Broglie wavelength equal to $10$ times its Compton wavelength in order to satisfy the assumption of nonrelativistic motion. Given these choices, all other parameters may be determined using the relations \er{Rparam}, \er{fixbeta}, and \er{fixR}.  
  
The zitterbewegung effect results in a downward shift in each dispersion curve in the present case, since the derivative of each eigenstate wave function at the boundary is opposite in sign to the boundary value of the wave function itself.  The spin-orbit splitting of the dispersion curve pairs $LP_{\abs{m_{\ell}} n^{\pm}}$ is exaggerated in the Figure by a factor of fifty for purposes of visualization.  Note that although we have included the dispersion curve for the $LP_{11}$ mode in our plot, since our perturbative treatment as formulated above does not apply to photons in this special case where $\abs{m_{\ell}}=1$ (see Appendix \ref{App:Elements}), we have plotted only a single uncorrected curve, $\plr{\beta_{0}}_{11}$ \textit{vs}. $R$.    

In the present simple case of a step profile for $\chi$, it is evident from equations \er{Rparam} above that the reduced photon wavelength at the waveguide center $\lambdabar_{\gamma}$ plays a role analogous to that of the electron Compton wavelength in defining a natural normalized frequency or waveguide parameter, which in turn determines the number of transversely bound states or guided modes present in the system.  More specifically, as the value for the waveguide radius $a$ approaches the value $\lambdabar_{C}$ for electrons (or $\lambdabar_{\gamma}$ for photons), thereby attempting to localize either particle within the guide, this localization is resisted as a larger percentage of each allowed transversely bound wave function penetrates into the region defined by $r>a$.  As a result of this, an increasing number of eigenstates are no longer guided by the confining potential or permittivity and thereby depart from the cylindrical guiding region (i.e., these modes become cut off).  In this way, $\lambdabar_{C}$ and $\lambdabar_{\gamma}$ are connected with the respective localizability of electrons and photons and by extension to various tunneling phenomena (cf. \cite{Keller99, Wang05}).  

The connection between $\lambdabar_{C}$ and $\lambdabar_{\gamma}$ and the strength of the zitterbewegung effect may be seen by noting that $\lambdabar_{dB}\approx\frac{c}{v}\lambdabar_{C}$ for nonrelativistic electrons, so that equations \er{beta0} yield $\plr{\beta_{0}^{e}}\approx \frac{v}{c}\frac{1}{\lambdabar_{C}}$ and $\plr{\beta_{0}^{\gamma}}\approx \frac{1}{\lambdabar_{\gamma}}$ for paraxially propagating particles obeying $\kappa_{0}\ll \beta_{0}$.  Substituting these relations into our effective Hamiltonian \er{Heff} yields the explicit result that the strength of the Darwin shift, and by extension the magnitude of the zitterbewegung, is proportional to the fundamental localizability scale $\lambdabar_{C}$ for electrons and $\lambdabar_{\gamma}$ for photons.

 \section{\label{sec:Conc}Conclusions\protect}

We have shown that when an electron or photon propagates in a cylindrically symmetric waveguide, it experiences both a zitterbewegung effect and a spin-orbit interaction leading to identical effective Hamiltonians, and therefore identical propagation dynamics, for both particles.  The presence of the zitterbewegung effect may be interpreted physically as a delocalization of the interaction between the particle and its confining influence, which in turn may be attributed to the the delocalization of the electron on the scale of its Compton wavelength, or the delocalization of the photon on the scale of its wavelength in the waveguide.  The analogy with the electron zitterbewegung, which has an amplitude on the order of the electron's reduced Compton wavelength, suggests that the magnitude of the photon zitterbewegung is of the order of the photon's reduced wavelength in the waveguide.  The presence of the spin-orbit interaction leads to the prediction of several rotational effects: the spatial or time evolution of either particle's spin/polarization vector is controlled by the sign of its orbital angular momentum quantum number, or conversely, its spatial wave function is controlled by its spin angular momentum.  Under either of these interactions, a two-state superposition of monoenergetic eigenstates with parallel and antiparallel spin and orbital angular momentum experiences a azimuthal precession of its Bloch vector in the relevant two-state space, with its direction controlled by the sign of the particle's spin or orbital angular momentum, and with a precession rate of $\abs{\delta\beta_{\text{SO}}}$ per unit $z$.  Each of these effects may occur as a function of either space (axial distance down the waveguide) or time.  We have argued elsewhere \cite{Leary09} that the common origin of the spin-orbit interaction for both particles is a geometric phase associated with the geometric evolution of either particle's spin vector as the particle propagates down the waveguide.  A more detailed geometric understanding of both the spin-orbit interaction and zitterbewegung effects may be attainable from an analysis of the first-order corrections to the particle wave functions under the action of the effective Hamiltonian given above.  Such an analysis will be the subject of a future paper.

\begin{acknowledgments}
We thank Michael Raymer for the foundational discussions that have led to this work.
 \end{acknowledgments}

\appendix

\section{\label{App:PhotonPerturbation}Photonic Perturbation Term\protect}

The purpose of this appendix is to express the photonic perturbation $\hat{H}^{\prime}\Psi=\nabla_{T}\bklr{\nabla_{T}\ln\epsilon\left(\rho\right)\cdot\Psi}$ in such a way that the spin-orbit interaction and zitterbewegung terms are explicitly manifest.  With this in view, we note that $\nabla_{T}\ln\epsilon\left(\rho\right)=\partial_{r}\ln\epsilon\left(\rho\right)\mathbf{\hat{r}}$ so that $\hat{H}^{\prime}\Psi$ simplifies to 

\bseq{Simplify}
\begin{align}
\nabla_{T}\left[\nabla_{T}\ln\epsilon\left(\rho\right)\cdot\Psi\right]&=\nabla_{T}\left(f\Psi_{r}\right) \lb{Simplifya} \\
&=\plr{\nabla_{T}f}\Psi_{r}+f\plr{\nabla_{T}\Psi_{r}} \lb{Simplifyb}
\end{align}
\eseq

\noindent where 

\beq{f}
	f\left(\rho\right)\equiv\partial_{r}\ln\epsilon\left(\rho\right)=\frac{-\Delta\chi^{\prime}}{a\plr{1-\Delta\chi}},
\eeq

\no and where $\Psi_{r}\left(\rho,\phi\right)\equiv\Psi\cdot\mathbf{\hat{r}}$ is the radial component of the transverse electric field $\Psi$ (recall that $\rho\eq\frac{r}{a}$). Employing a circularly polarized basis with basis vectors $\unit{e}_{\pm} \eq\frac{1}{\sqrt{2}}\plr{\unit{x}\pm i\unit{y}}$, we may express the transverse gradient operator as 

\beq{gradT}
\nabla_{T}=\hat{\mathbf{e}}_{+}\nabla_{+}+\hat{\mathbf{e}}_{-}\nabla_{-}
\eeq

\no with 

\beq{gradsub}
\nabla_{\pm}\equiv \frac{1}{\sqrt{2}}e^{\mp i\phi}\left(\partial_{r}\mp\frac{i}{r}\partial_{\phi}\right),
\eeq

\no while the radial component of the electric field may be written as 

\beq{rad}
\Psi_{r}=\frac{1}{\sqrt{2}}\left(e^{+i\phi}\Psi_{+}+e^{-i\phi}\Psi_{-}\right),
\eeq

\no with $\Psi_{\pm}\equiv\Psi\cdot\hat{\mathbf{e}}_{\pm}^{*}$. Substituting these relations into the right-hand side of \eqref{Simplifyb}, using $\nabla_{\pm}f=\frac{1}{\sqrt{2}}e^{\mp i\phi}\plr{\pd{r}f}$, and employing the following column vector notation for  $\Psi$,

\beq{Column}
\Psi=\Psi_{+}\hat{\mathbf{e}}_{+}+\Psi_{-}\hat{\mathbf{e}}_{-}\equiv\left(\begin{array}{c}{\Psi_{+}}\\{\Psi_{-}}\end{array}\right),
\eeq

\no readily yields the first line of the following succession of expressions:

\begin{widetext}
\begin{subequations}
\begin{align}
\hat{H}^{\prime}\Psi &= \frac{1}{2}
\left[\begin{array}{cc} {f^{\prime}+fe^{-i\phi}\left(\partial_{r}-\frac{i}{r}\partial_{\phi}\right)e^{+i\phi}} & {f^{\prime}e^{-2i\phi}+fe^{-i\phi}\left(\partial_{r}-\frac{i}{r}\partial_{\phi}\right)e^{-i\phi}} \\ {f^{\prime}e^{+2i\phi}+fe^{+i\phi}\left(\partial_{r}+\frac{i}{r}\partial_{\phi}\right)e^{+i\phi}} & {f^{\prime}+fe^{+i\phi}\left(\partial_{r}+\frac{i}{r}\partial_{\phi}\right)e^{-i\phi}}\end{array}\right]\left(\begin{array}{c}{\Psi_{+}}\\{\Psi_{-}}\end{array}\right), \label{Matrix1} \\[2mm]
& =\Bigg\{\frac{1}{2}\Bigg(\plr{f^{\prime}+f\partial_{r}}\mathbf{1}+\frac{f}{r}\hat{\sigma}_{z}\mathbf{Q}\hat{\ell}_{z}\mathbf{Q}\Bigg)+
\frac{1}{2}\Bigg(\plr{f^{\prime}+f\partial_{r}}\mathbf{\hat{N}^{\plr{\gamma}}}+\frac{f}{r}\hat{\sigma}_{z}\mathbf{Q}\hat{\sigma}_{x}\hat{\ell}_{z}\mathbf{Q}\Bigg)\Bigg\}\left(\begin{array}{c}{\Psi_{+}}\\{\Psi_{-}}  \end{array}\right), \label{Matrix2} \\[2mm]
&=\Bigg\{\frac{1}{2}\Bigg[\plr{f^{\prime}+f\partial_{r}+\frac{f}{r}}+\frac{f}{r}\hat{\sigma}_{z}\hat{\ell}_{z}\Bigg]\mathbf{1}+\frac{1}{2}\Bigg[\plr{f^{\prime}+f\partial_{r}+\frac{f}{r}}+\frac{f}{r}\hat{\sigma}_{z}\hat{\ell}_{z}\Bigg]\mathbf{\hat{N}^{\plr{\gamma}}}\Bigg\}\left(\begin{array}{c}{\Psi_{+}}\\{\Psi_{-}}  \end{array}\right). \label{Matrix3}, \\[2mm]
&=\plr{\hat{H}_{\text{Dar}}^{\plr{\gamma}}+\hat{H}_{\text{SO}}^{\plr{\gamma}}}\plr{\mathbf{1}+\mathbf{\hat{N}^{\plr{\gamma}}}}\Psi \label{Matrix4}
\end{align}
\end{subequations}
\end{widetext}

In \er{Matrix2}--\er{Matrix4}, $\hat{\sigma}_x$ and $\hat{\sigma}_{z}$ are respectively the Pauli $x$ and $z$ matrices, the primes on the $f$ functions denote derivatives with respect to $r\eq\rho a$, and 

\bseq{Matrices}
\begin{align}
\mathbf{Q}&\equiv\left(\begin{array}{cc}{0}&{e^{-i\phi}}\\{e^{+i\phi}}&{0}\end{array}\right), \lb{Matricesa} \\
\mathbf{\mathbf{\hat{N}^{\plr{\gamma}}}}&\equiv\mathbf{Q}\hat{\sigma}_{x}\mathbf{Q}=\left(\begin{array}{cc}{0}&{e^{-2i\phi}}\\{e^{+2i\phi}}&{0}\end{array}\right), \lb{Matricesb} \\
&\;\;\;\;\;\;\;\;\;\;\;\;\;\;\;\eq\hat{\sigma}^{2}_{+}\hat{\ell}^{2}_{-}+\hat{\sigma}^{2}_{-}\hat{\ell}^{2}_{+},
\end{align}
\eseq

\no while  $\hat{H}_{\text{Dar}}^{\plr{\gamma}}$ and $\hat{H}_{\text{SO}}^{\plr{\gamma}}$ are defined in \er{PPert}. In deriving \er{Matrix3} from \er{Matrix2}, we have made use of the operator identities $\hat{\ell}_{z}\mathbf{Q}=\mathbf{Q}\hat{\sigma}_{z}+\mathbf{Q}\hat{\ell}_{z}$, $\mathbf{Q}^{2}=\hat{\sigma}_{z}^{2}=\mathbf{1}$, $\hat{\sg}_x \hat{\ell}_z =\hat{\ell}_z \hat{\sg}_x$, $\mbf{Q}\hat{\sg}_z =-\hat{\sg}_z \mbf{Q}$, and $\hat{\sigma}_{z}\mathbf{\hat{N}^{\plr{\gamma}}}=-\mathbf{\hat{N}^{\plr{\gamma}}}\hat{\sigma}_{z}$ in order to simplify the matrix products in \er{Matrix2}, yielding the results

\begin{align} \label{Diagonal}
\hat{\sigma}_{z}\mathbf{Q}\hat{\ell}_{z}\mathbf{Q} & =\mathbf{1}+\hat{\sigma}_{z}\hat{\ell}_{z} \nonumber \\
\hat{\sigma}_{z}\mathbf{Q}\hat{\sigma}_{x}\hat{\ell}_{z}\mathbf{Q} & =\left(\mathbf{1}+\hat{\sigma}_{z}\hat{\ell}_{z}\right)\mathbf{N}.
\end{align}

\no From \er{Matrix3} it is clear that $\hat{H}^{\prime}\Psi$ factors into the result \er{Matrix4}, where $f^{\prime}\eq\pd{r}f=\frac{-1}{\plr{1-\Delta\chi}}\frac{\Delta}{a^2}\plr{\chi^{\prime\prime}+\frac{\Delta\plr{\chi^{\prime}}^2 }{1-\Delta\chi}}$ from \er{f}.  This is equation \er{PHPrime} in the main body of this work.

\section{\label{App:Elements}Perturbation Matrix Elements\protect}

In this appendix we use perturbation theory to calculate the matrix elements of the electronic and photonic perturbation Hamiltonians given in \er{EHPrime} and \er{PHPrime}, and the corresponding corrections to the propagation constant $\plr{\beta_{0}^{2}}_{n\abs{m_{\ell}}}$, in the weakly guided regime where $\Delta\ll 1$.  For electrons, straightforward application of \er{innerproduct} to the Hamiltonian in \er{EHPrime} yields

\begin{align} \label{ApplE}
&\bra{n^{\prime}\, m_{\ell}^{\prime}\, \sigma^{\prime}}\hat{H}^{\prime}\ket{n\, m_{\ell}\, \sigma}= \non \\
&\;\;\;\;\;\frac{2\pi}{\lambdabar_{C}^2}\left<\plr{\frac{\hbar\omega}{mc^2}-\Delta\chi}^2\right>_{}\,\delta_{m_{\ell }^{\prime} m_{\ell }} \delta_{\sg^{\prime} \sg} \non \\
&\;-\frac{\pi\Delta}{a^2}\Bigg[\!\plr{\frac{1}{2}\left< \chi^{\prime\prime} + \frac{\chi^{\prime}}{\rho}\right> +\sg m_{\ell }\left< \frac{\chi^{\prime} }{\rho}\right>}\!\delta_{m_{\ell }^{\prime} m_{\ell }} \delta_{\sg^{\prime} \sg} \Bigg. \non \\
& \;\; \Bigg.+i\plr{\beta_{0}}_{n\abs{m_{\ell}}}\!\bigg<\!\chi^{\prime} \!\bigg>\!\plr{\!\delta_{m_{\ell }^{\prime}  m_{\ell }-1} \, \delta_{\sg^{\prime} \sg+1}-\delta_{m_{\ell }^{\prime} m_{\ell }+1} \, \delta_{\sg^{\prime} \sg-1}\!}\!\Bigg],
\end{align}

\no where the bracket notation $\left<\hat{O}\right>$ denotes the radial integral $\left< \hat{O}\right> \eq \int{\psi_{n^{\prime}\abs{m_{\ell}^{\prime}}}\, \hat{O} \; \psi_{n\abs{m_{\ell}}} \rho d\rho} $ for any operator $\hat{O}$ contained within.  

For photons, we begin by noting that to first order in $\Delta$, the final term in \er{PPerta} is negligible, while the prefactor  $\frac{1}{1-\Delta\chi\plr{\rho}}$ in \er{PPerta} and \er{PPertb} is approximately equal to unity, so that

\bseq{1stOrderPhoton}
\begin{align}
\hat{H}_{\text{Dar}}^{\plr{\gamma}}&\approx-\frac{\Delta}{2a^2}\plr{\chi^{\prime\prime} +\frac{\chi' }{\rho}+\chi^{\prime} \pd{\rho}}, \lb{1stOrderPhotona} \\
\hat{H}_{\text{SO}}^{\plr{\gamma}}&\approx-\frac{\Delta}{2a^2}\frac{\chi^{\prime}}{\rho}\hat{\sigma}_{z} \hat{\ell}_{z}. \lb{1stOrderPhotonb}
\end{align}
\eseq

\no Substituting \er{1stOrderPhotona} and \er{1stOrderPhotonb} into \er{PHPrime} and employing \er{innerproduct}, we find that for photons,

\begin{align} \label{ApplP}
& \bra{n^{\prime}\, m_{\ell}^{\prime}\, \sigma^{\prime}}\hat{H}^{\prime}\ket{n\, m_{\ell}\, \sigma} = -\frac{\pi\Delta}{a^2} \;\, \times \non \\
&\Bigg[\plr{\left<\!\chi^{\prime\prime}+\frac{\chi^{\prime}}{\rho}+\chi^{\prime} \pd{\rho} \!\right>+\sg m_{\ell }\left< \frac{\chi^{\prime} }{\rho}\right>}\delta_{m_{\ell }^{\prime} m_{\ell }} \delta_{\sg^{\prime} \sg} \Bigg. \non \\
&+\! \left<\!\chi^{\prime\prime}+\frac{\chi^{\prime}}{\rho}+\chi^{\prime} \pd{\rho} \!\right>\!\plr{\delta_{m_{\ell }^{\prime} m_{\ell }-2} \, \delta_{\sg^{\prime} \sg+2}+\delta_{m_{\ell }^{\prime} \, m_{\ell }+2} \delta_{\sg^{\prime} \sg-2}} \Bigg. \non \\
&+\Bigg. \left<\frac{\chi^{\prime} }{\rho}\right>\bigg(\plr{m_{\ell }-2}\delta_{m_{\ell }^{\prime} m_{\ell }-2} \, \delta_{\sg^{\prime}\sg+2}\bigg. \non \\
&\;\;\;\;\;\;\;\;\;\;\;\;\;\,- \bigg.\plr{m_{\ell }+2}\delta_{m_{\ell }^{\prime} m_{\ell }+2} \, \delta_{\sg^{\prime} \sg-2}\bigg) \Bigg],
\end{align}

\no where we have made the replacements $\plr{\sg+2}\to+1$ and $\plr{\sg-2}\to-1$ in the last two lines of \er{ApplP} since the Kronecker delta functions $\delta_{\sg^{\prime} \, \sg+2}$ and $\delta_{\sg^{\prime} \, \sg-2}$ act as selection rules, respectively forcing $\sg=-1$ and $\sg=+1$.

It follows from \er{ApplE} and \er{ApplP}, that the explicit forms for the perturbation matrix $\bra{n^{\prime}\, m_{\ell}^{\prime}\, \sigma^{\prime}}\hat{H}'\ket{n\, m_{\ell}\, \sigma}$ are:

\begin{widetext}
\footnotesize{
\[
  \begin{blockarray}{c cc cccc cccc cccc cc cccc c}
&  
&  
& {\text{\begin{sideways}$\ket{n\; 0\, +}$\end{sideways}}}  
&{\text{\begin{sideways}$\ket{n\; 0\, -}$\end{sideways}}}
&{\text{\begin{sideways}$\ket{n\; {+1}\, -}$\end{sideways}}}
&{\text{\begin{sideways}$\ket{n\; {-\!1}\, +}$\end{sideways}}}
&{\text{\begin{sideways}$\ket{n\; {+1}\, +}$\end{sideways}}}
&{\text{\begin{sideways}$\ket{n\;\, {-\!1}\, -}$\end{sideways}}}
&{\text{\begin{sideways}$\ket{n\; {+2}\, -}$\end{sideways}}}
&{\text{\begin{sideways}$\ket{n\; {-\!2}\, +}$\end{sideways}}}
&{\text{\begin{sideways}$\ket{n\; {+2}\, +}$\end{sideways}}}
&{\text{\begin{sideways}$\ket{n\; {-\!2}\, -}$\end{sideways}}}
&{\text{\begin{sideways}$\ket{n\; {+3}\, -}$\end{sideways}}}
&{\text{\begin{sideways}$\ket{n\; {-\!3}\, +}$\end{sideways}}}
&{\text{\begin{sideways}$\ket{n\; {+3}\, +}$\end{sideways}}}
&{\text{\begin{sideways}$\ket{n\; {-\!3}\, -}$\end{sideways}}}
&{\text{\begin{sideways}$\ket{n\; {+4}\, -}$\end{sideways}}}
&{\text{\begin{sideways}$\ket{n\; {-\!4}\, +}$\end{sideways}}}
&{\text{\begin{sideways}$\ket{n\; {+4}\, +}$\end{sideways}}}
&{\text{\begin{sideways}$\ket{n\; {-\!4}\, -}$\end{sideways}}}
&{\cdots} 
& $\;$ \\ 
    \begin{block}{				r(@{\hspace*{-4pt} \;\;\;\;\; }	c c	cc  |	cccc	 |	cccc  |	cccc  |	cccc  |	c)@{\hspace*{-4pt}}}
{\bra{n^{\prime}\, 0\, +}} 		&&&A_{0}&&-iB_{1}&&&&&&&&&&&&&&&&\\
{\bra{n^{\prime}\, 0\, -}}   		&&&&A_{0}&&+iB_{1}&&&&&&&&&&&&&&&\\\cline{4-23}
{\bra{n^{\prime}\, {+1}\, -}} 		&&&+iB_{0}&&A_{1}^{+}&&&&&&&&&&&&&&&&\\
{\bra{n^{\prime}\, {-\!1}\,+}} 	&&&&-iB_{0}&&A_{1}^{+}&&&&&&&&&&&&&&&\\
{\bra{n^{\prime}\, {+1}\, +}} 	&&&&&&&A_{1}^{-}&&-iB_{2}&&&&&&&&&&&&\\
{\bra{n^{\prime}\, {-\!1}\, -}} 	&&&&&&&&A_{1}^{-}&&+iB_{2}&&&&&&&&&&&\\\cline{4-23}
{\bra{n^{\prime}\, {+2}\, -}} 		&&&&&&&+iB_{1}&&A_{2}^{+}&&&&&&&&&&&&\\
{\bra{n^{\prime}\, {-\!2}\, +}}	&&&&&&&&-iB_{1}&&A_{2}^{+}&&&&&&&&&&&\\
{\bra{n^{\prime}\, {+2}\, +}}  	&&&&&&&&&&&A_{2}^{-}&&-iB_{3}&&&&&&&&\\
{\bra{n^{\prime}\, {-\!2}\, -}} 	&&&&&&&&&&&&A_{2}^{-}&&+iB_{3}&&&&&&&\\\cline{4-23}
{\bra{n^{\prime}\, {+3}\, -}}  	&&&&&&&&&&&+iB_{2}&&A_{3}^{+}&&&&&&&&\\
{\bra{n^{\prime}\, {-\!3}\, +}} 	&&&&&&&&&&&&-iB_{2}&&A_{3}^{+}&&&&&&&\\
{\bra{n^{\prime}\, {+3}\, +}} 	&&&&&&&&&&&&&&&A_{3}^{-}&&-iB_{4}&&&&\\
{\bra{n^{\prime}\, {-\!3}\, -}}  	&&&&&&&&&&&&&&&&A_{3}^{-}&&+iB_{4}&&&\\\cline{4-23}
{\bra{n^{\prime}\, {+4}\, -}}  	&&&&&&&&&&&&&&&+iB_{3}&&A_{4}^{+}&&&&\\
{\bra{n^{\prime}\, {-\!4}\, +}}	&&&&&&&&&&&&&&&&-iB_{3}&&A_{4}^{+}&&&\\
{\bra{n^{\prime}\, {+4}\, +}}  	&&&&&&&&&&&&&&&&&&&A_{4}^{-}&&\\
{\bra{n^{\prime}\, {-\!4}\, -}}  	&&&&&&&&&&&&&&&&&&&&A_{4}^{-}&\dots\\\cline{4-23}
&&& \vdots &&&&&&&&&&&&&&&&&\vdots&\ddots\\
    \end{block}
  \end{blockarray}
\]}

\normalsize
\vspace{4mm}
\bseq{Efactors}
\begin{align}
A_{\abs{m_{\ell}}}^{\pm}&\eq\frac{2 \pi}{\lambdabar_{C}^2} \left<\plr{\frac{\hbar\omega}{mc^2}-\Delta\chi}^2\right>+\frac{\pi\Delta}{a^2}\plr{-\frac{1}{2}\bigg< \chi^{\prime\prime} + \frac{\chi^{\prime}}{\rho} \bigg> \pm \abs{m_{\ell }}\left< \frac{\chi' }{\rho}\right>}, \lb{Efactorsa}  \\
B_{\abs{m_{\ell}}}&\eq\frac{\pi\Delta}{a^2}\plr{\beta_{0}}_{n\abs{m_{\ell}}}\bigg<\chi' \bigg>, \;\;\;\;\;\;\;\;\;\; \;\;\;\;\;\;\;\;\;\;  \;\;\;\;\;\;\;\;\;\;  \;\;\;\;\;\;\;\;\;\;  \text{\textbf{(for electrons)}} \lb{Efactorsb}
\end{align}
\eseq

\vspace{5mm}

\footnotesize{
\[
  \begin{blockarray}{c cc cccc cccc cccc cc cccc c}
&  
&  
& {\text{\begin{sideways}$\ket{n\; 0\, +}$\end{sideways}}}  
&{\text{\begin{sideways}$\ket{n\; 0\, -}$\end{sideways}}}
&{\text{\begin{sideways}$\ket{n\; {+1}\, -}$\end{sideways}}}
&{\text{\begin{sideways}$\ket{n\; {-\!1}\, +}$\end{sideways}}}
&{\text{\begin{sideways}$\ket{n\; {+1}\, +}$\end{sideways}}}
&{\text{\begin{sideways}$\ket{n\;\, {-\!1}\, -}$\end{sideways}}}
&{\text{\begin{sideways}$\ket{n\; {+2}\, -}$\end{sideways}}}
&{\text{\begin{sideways}$\ket{n\; {-\!2}\, +}$\end{sideways}}}
&{\text{\begin{sideways}$\ket{n\; {+2}\, +}$\end{sideways}}}
&{\text{\begin{sideways}$\ket{n\; {-\!2}\, -}$\end{sideways}}}
&{\text{\begin{sideways}$\ket{n\; {+3}\, -}$\end{sideways}}}
&{\text{\begin{sideways}$\ket{n\; {-\!3}\, +}$\end{sideways}}}
&{\text{\begin{sideways}$\ket{n\; {+3}\, +}$\end{sideways}}}
&{\text{\begin{sideways}$\ket{n\; {-\!3}\, -}$\end{sideways}}}
&{\text{\begin{sideways}$\ket{n\; {+4}\, -}$\end{sideways}}}
&{\text{\begin{sideways}$\ket{n\; {-\!4}\, +}$\end{sideways}}}
&{\text{\begin{sideways}$\ket{n\; {+4}\, +}$\end{sideways}}}
&{\text{\begin{sideways}$\ket{n\; {-\!4}\, -}$\end{sideways}}}
&{\cdots} 
& $\;$ \\ 
    \begin{block}{r(@{\hspace*{-15pt } \;\;\;\;\;\;\;\;\;\;}cccc|cccc|cccc|cccc|cccc|c)@{\hspace*{-4pt}}
}
{\bra{n^{\prime}\, 0\, +}} 		&&&A_{0}&&&&&&B_{2}^{-}&&&&&&&&&&&&\\
{\bra{n^{\prime}\, 0\, -}} 		&&&&A_{0}&&&&&&B_{2}^{-}&&&&&&&&&&&\\\cline{4-23}
{\bra{n^{\prime}\, {+1}\, -}} 	&&&&&A_{1}^{+}&A_{1}^{+}&&&&&&&&&&&&&&&\\
{\bra{n^{\prime}\, {-\!1}\,+}} 	&&&&&A_{1}^{+}&A_{1}^{+}&&&&&&&&&&&&&&&\\
{\bra{n^{\prime}\, {+1}\, +}} 	&&&&&&&A_{1}^{-}&&&&&&B_{3}^{-}&&&&&&&&\\
{\bra{n^{\prime}\, {-\!1}\, -}}  	&&&&&&&&A_{1}^{-}&&&&&&B_{3}^{-}&&&&&&&\\\cline{4-23}
{\bra{n^{\prime}\, {+2}\, -}}  	&&&B_{0}^{+}&&&&&&A_{2}^{+}&&&&&&&&&&&&\\
{\bra{n^{\prime}\, {-\!2}\, +}}	&&&&B_{0}^{+}&&&&&&A_{2}^{+}&&&&&&&&&&&\\
{\bra{n^{\prime}\, {+2}\, +}} 	&&&&&&&&&&&A_{2}^{-}&&&&&&B_{4}^{-}&&&&\\
{\bra{n^{\prime}\, {-\!2}\, -}} 	&&&&&&&&&&&&A_{2}^{-}&&&&&&B_{4}^{-}&&&\\\cline{4-23}
{\bra{n^{\prime}\, {+3}\, -}}     &&&&&&&B_{1}^{+}&&&&&&A_{3}^{+}&&&&&&&&\\
{\bra{n^{\prime}\, {-\!3}\, +}}  	&&&&&&&&B_{1}^{+}&&&&&&A_{3}^{+}&&&&&&&\\
{\bra{n^{\prime}\, {+3}\, +}} 	&&&&&&&&&&&&&&&A_{3}^{-}&&&&&&\\
{\bra{n^{\prime}\, {-\!3}\, -}} 	&&&&&&&&&&&&&&&&A_{3}^{-}&&&&&\dots\\\cline{4-23}
{\bra{n^{\prime}\, {+4}\, -}}	&&&&&&&&&&&B_{2}^{+}&&&&&&A_{4}^{+}&&&&\\
{\bra{n^{\prime}\, {-\!4}\, +}} 	&&&&&&&&&&&&B_{2}^{+}&&&&&&A_{4}^{+}&&&\\
{\bra{n^{\prime}\, {+4}\, +}}  	&&&&&&&&&&&&&&&&&&&A_{4}^{-}&&\\
{\bra{n^{\prime}\, {-\!4}\, -}} 	&&&&&&&&&&&&&&&&&&&&A_{4}^{-}&\\\cline{4-23}
&&& \vdots &&&&&&&&&&&&&\vdots&&&&&\ddots\\
    \end{block}
  \end{blockarray}
\]}
\normalsize
\vspace{4mm}
\bseq{Pfactors}
\begin{align}
A_{m_{\ell}}^{\pm}&\eq\frac{\pi\Delta}{a^2}\plr{-\left<\!\chi^{\prime\prime}+\frac{\chi^{\prime}}{\rho}+\chi^{\prime} \pd{\rho} \!\right> \pm m_{\ell }\left< \frac{\chi^{\prime} }{\rho}\right>}, \lb{Pfactorsa}  \\
B_{m_{\ell}}^{\pm}&\eq\frac{\pi\Delta}{a^2}\plr{-\left<\!\chi^{\prime\prime}+\frac{\chi^{\prime}}{\rho}+\chi^{\prime} \pd{\rho} \!\right>\pm\plr{m_{\ell}\pm 2}\left< \frac{\chi^{\prime} }{\rho}\right>}, 
\;\;\;\;\;\;\;\;\;\;  \text{\textbf{(for photons)}} \lb{Pfactorsb} 
\end{align}
\eseq
\end{widetext}

\no For each allowed value of $n$ and $\abs{m_{\ell}}$, the unperturbed squared propagation constant $\plr{\beta_{0}^{2}}_{n\abs{m_{\ell}}}$ is degenerate in the subspace spanned by the states $\{\ket{n\, \abs{m_{\ell}}\, +},\;\ket{n\, -\!\abs{m_{\ell}}\, -},\;\ket{n\, \abs{m_{\ell}}\, -},\;\ket{n\, -\!\abs{m_{\ell}}\, +}\}$, which is four dimensional except in the case where $\abs{m_{\ell}}=0$.  According to first-order degenerate perturbation theory \cite{McIntyre}, we therefore need to diagonalize the submatrices $\bra{n\, m_{\ell}'\, \sigma'}\hat{H}'\ket{n\, m_{\ell}\, \sigma}$ of the perturbation matrix within each such degenerate subspace where $|m_{\ell}'|=|m_{\ell}|$ in order to find the first-order shifts $\delta\!\plr{\beta_{0}^{2}}_{n\abs{m_{\ell}}}$ due to the perturbation.  The vertical and horizontal lines in the above matrices delineate these degenerate subspaces.  It is clear by inspection that the appropriate subspaces are already diagonal in the $\ket{n\, m_{\ell}\, \sigma}$ eigenbasis, with the single exception of $\abs{m_{\ell}}=1$ in the photon case \footnote{Because of this, for photons we exclude the states where $\abs{m_{\ell}}=1$ from the subsequent discussion.}.   Since our unperturbed states $\ket{n\, m_{\ell}\, \sigma}$ are therefore a valid basis for the application of degenerate perturbation theory, we may readily calculate the first-order shifts in $\plr{\beta_{0}^{2}}_{n\abs{m_{\ell}}}$ for both particles by setting $n^{\prime}=n$, $m_{\ell}^{\prime}=m_{\ell}$, and $\sg^{\prime}=\sg$ in \er{ApplE} and \er{ApplP}, so that only the diagonal terms proportional to $\delta_{m_{\ell }' m_{\ell }} \delta_{\sg' \sg}$ contribute to $\delta\!\plr{\beta_{0}^{2}}_{n\abs{m_{\ell}}}$:

\bseq{ApplicationBoth}
\begin{align} 
\delta\!\plr{\beta_{0}^{2}}_{n\abs{m_{\ell}}}=&\bra{n\, m_{\ell}\, \sigma}\hat{H}^{\prime}\ket{n\, m_{\ell}\, \sigma} \non \\
=&\frac{2\pi}{\lambdabar_{C}^2}\left<\plr{\frac{\hbar\omega}{mc^2}-\Delta\chi}^2\right>_{}\, 
\non \\
&-\frac{\pi\Delta}{a^2}\plr{\frac{1}{2}\left< \chi^{\prime\prime} + \frac{\chi^{\prime}}{\rho}\right> +\sg m_{\ell }\left< \frac{\chi^{\prime} }{\rho}\right>}  \non \\
& \!\!\!\!\!\!\!\!\!\! \!\!\!\!\!\!\!\!\!\! \!\!\!\!\!\!\!\!\!\! \!\!\!\!\!\!\!\!\!\! \text{\textbf{$ \;\;\; $ (for electrons)}} \lb{ApplicationBothA}  \\
\non \\
\delta\!\plr{\beta_{0}^{2}}_{n\abs{m_{\ell}}}=&\bra{n\, m_{\ell}\, \sigma}\hat{H}^{\prime}\ket{n\, m_{\ell}\, \sigma} \non \\
&=-\frac{\pi\Delta}{a^2}\left<\!\chi^{\prime\prime}+\frac{\chi^{\prime}}{\rho}+\chi^{\prime} \pd{\rho} \!\right>+\sg m_{\ell }\left< \frac{\chi^{\prime} }{\rho}\right> \non \\
& \!\!\!\!\!\!\!\!\!\! \!\!\!\!\!\!\!\!\!\! \!\!\!\!\!\!\!\!\!\! \!\!\!\!\!\!\!\!\!\! \text{\textbf{$ \;\;\; $ (for photons)}} \lb{ApplicationBothB} 
\end{align}
\eseq

Considering now simultaneously the ``Darwin term'' inner products $\frac{1}{2}\left< \chi^{\prime\prime} + \frac{\chi^{\prime}}{\rho}\right>$ and $\left<\!\chi^{\prime\prime}+\frac{\chi^{\prime} }{\rho}+\chi^{\prime} \pd{\rho} \!\right>$ in \er{ApplicationBothA} and \er{ApplicationBothB}, we integrate the terms involving $\chi^{\prime\prime}$ by parts for each, the result of which cancels the remaining terms, resulting in the same expression for both cases:

\begin{align}\lb{Dar}
\frac{1}{2}\left< \chi^{\prime\prime} + \frac{\chi^{\prime}}{\rho}\right>&=\left<\chi^{\prime\prime}+\frac{\chi^{\prime} }{\rho}+\chi^{\prime} \pd{\rho} \right> \non \\
&=-\int{\!\chi^{\prime} \psi_{n \abs{m_{\ell}}}\pd{\rho}\psi_{n\abs{m_{\ell}}} \rho \, d\rho} \non \\
&\eq-\bigg<\chi^{\prime}\pd{\rho}\bigg>_{n\abs{m_{\ell}}}.
\end{align}

\no where $\left<\hat{O}\right>_{\!\!n \abs{m_{\ell}}}$ denotes the radial integral $\left<\hat{O}\right>$ with $n^{\prime}=n$ and $m_{\ell}^{\prime}=m_{\ell}$, for any operator $\hat{O}$ contained within.  Substitution of \er{Dar} into \er{ApplicationBothA} and \er{ApplicationBothB} immediately results in equation \er{deg} in the main body of this paper.

We note here that the perturbation $\hat{H}^{\prime}$ is not Hermitian with respect to the inner product given in \er{innerproduct}, as is evident by inspection of equations \er{ApplE} and \er{ApplP} or their corresponding explicit matrices.  However, the \emph{unperturbed} Hamiltonian operator $\hat{H}_{0}$ \emph{is} Hermitian for either choice of $k\plr{\rho}$ given in \er{ksqelec} and \er{ksqphot}, according to a Sturm--Liouville analysis of the unperturbed equation \er{unpert}, so that the unperturbed eigenstates $\ket{n\, m_{\ell}\, \sigma}$ form a complete set and have real eigenvalues $\plr{\beta_{0}^{2}}_{n \, \abs{m_{\ell}}}$.  The standard results in perturbation theory for the first-order corrections to the eigenvalues $\plr{\beta_{0}^{2}}_{n\abs{m_{\ell}}}$ and eigenstates $\ket{n\, m_{\ell}\, \sigma}$ of $\hat{H}_{0}$ depend only on the assumption of the Hermiticity of $\hat{H}_{0}$, so that the first--order corrections $\delta\!\plr{\beta_{0}^{2}}_{n\abs{m_{\ell}}}$ given in \er{deg} are valid even though $\hat{H}^{\prime}$ is not Hermitian \footnote{Similar conclusions have been verified for multiple refractive index profiles in \cite{Liberman92} and \cite{Volyar98} in the context of the spin-orbit interaction of photons in optical fibers.}.  However, the standard formula for the second--order eigenvalue corrections $\delta^{\plr{2}}\!\plr{\beta_{0}^{2}}_{n\abs{m_{\ell}}}$ does undergo a slight modification for a non-Hermitian perturbation: 

\begin{align} \lb{2ndorder}
&\delta^{\plr{2}}\!\plr{\beta_{0}^{2}}_{n\abs{m_{\ell}}}= \non \\
&\;\; \sum_{n^{\prime}\neq n}\frac{\bra{n^{\prime}\, m_{\ell}\, \sigma}H^{\prime}\ket{n\, m_{\ell}\, \sigma}\bra{n\, m_{\ell}\, \sigma}H^{\prime}\ket{n^{\prime}\, m_{\ell}\, \sigma}}{\plr{\beta_{0}^{2}}_{n\abs{m_{\ell}}}-\plr{\beta_{0}^{2}}_{n^{\prime}\abs{m_{\ell}}}}
\end{align}

\no All corrections to $\plr{\beta_{0}^{2}}_{n\abs{m_{\ell}}}$ are real for the perturbation Hamiltonians considered in this paper, as is to be expected.

\end{document}